Technical paper

# Illustrating the benefits of efficient creation and adaption of behavior models in intelligent Digital Twins over the machine life cycle

Daniel Dittler [a,*,1], Valentin Stegmaier [a,b,c,1], Nasser Jazdi [a,2], Michael Weyrich [a,3]

[a] Institute of Industrial Automation and Software Engineering (IAS), University of Stuttgart, Pfaffenwaldring 47, Stuttgart 70569, Germany
[b] Graduate School of Excellence advanced Manufacturing Engineering (GSaME), University of Stuttgart, Nobelstraße 12, Stuttgart 70569, Germany
[c] J. Schmalz GmbH, Johannes-Schmalz-Straße 1, Glatten 72293, Germany

ABSTRACT

The concept of the Digital Twin, which in the context of this paper is the virtual representation of a production system or its components, can be used as a "digital playground" to master the increasing complexity of these assets. One of the central subcomponents of the Digital Twin are behavior models that can enable benefits over the entire machine life cycle. However, the creation, adaption and use of behavior models throughout the machine life cycle is very time-consuming, which is why approaches to improve the cost-benefit ratio are needed. Furthermore, there is a lack of specific use cases that illustrate the application and added benefit of behavior models over the machine life cycle, which is why the universal application of behavior models in industry is still lacking compared to research. This paper first presents the fundamentals, challenges and related work on Digital Twins and behavior models in the context of the machine life cycle. Then, concepts for low-effort creation and automatic adaption of Digital Twins are presented, with a focus on behavior models. Finally, the aforementioned gap between research and industry is addressed by demonstrating various realized use cases over the machine life cycle, in which the advantages as well as the application of behavior models in the different life cycle phases are shown.

## 1. Introduction

Automation in industry is experiencing increasing modularization and digitalization. The ubiquitous availability of data and services opens up new perspectives such as the vision of self-organized, adaptive and partially self-organizing flexible production systems for ever shorter product life cycles and an increasing product variance [1,2]. As a result, production can be made more efficient and sustainable, e.g. through shorter setup times and the optimized use of energy and resources [3]. With these trends, production systems are becoming more and more complex, as they consist of a multitude of different individual components with links and dependencies among each other. In order to make the complexity more manageable, a Digital Twin (DT), which serves as a virtual representation of the physical asset, is becoming increasingly important [4]. A central component of these DTs are models that abstract the production system from different perspectives [5]. Some of these models are used to shorten development times, for virtual commissioning, operation parallel simulation and many more use cases. Behavior models, commonly known as simulation models, are used for this purpose. Mostly, physics-based models (White-box models) are applied to test components in advance in the virtual space. Data-driven behavior models (Black-box models) or the combination of White and Black-box equal to Grey-box can be used in some use cases. Nonetheless, behavior models as White-boxes have the benefit that they are generalizable to problems of similar physics and less susceptible to bias [6]. Therefore, the authors consider White-box behavior models especially suitable for DTs. However, the creation, adaption and use of the models in particular is very time-consuming, which is why industrial practice is still far away from the universal use of behavior models across the machine life cycle [7]. A possible solution to this can be the automated creation, adaption and usage of behavior models in industrial automation improving the effort-benefit ratio using such models. By automating the mentioned activities arising from the actual automation of industrial processes, one can also speak of the automation of automation.

* Corresponding author.
*E-mail address:* daniel.dittler@ias.uni-stuttgart.de (D. Dittler).
[1] These authors contributed equally to the work.
[2] Writing – Review & Editing
[3] Project Administration

From the authors' point of view, the concept of DTs is a possible solution to the challenges mentioned above, but it requires novel approaches to avoid shifting effort from the real to the virtual world, but to add real value. There is still a need for action, especially for the low-cost creation and adaptation of behavior models in the DT. Furthermore, from the authors' point of view, the literature often describes only conceptual use cases of behavior models over the machine life cycle. However, there is a lack of specific implementation of machine life cycle use cases, which is why there is a gap in the use of behavior models between research and industry.

Based on these findings, this paper has two main objectives. The first objective of this paper is to present and combine the concepts of automatic creation and adaption of DTs with a focus on behavior models. This simplifies the consistent use of the behavior models from the DT along the entire machine lifecycle and across company boundaries. The second objective of this paper is to illustrate the application and benefits of behavior models created and adapted using the presented concepts along the entire machine life cycle by means of exemplary use cases to address the mentioned gap.

The remainder of this paper is organized as follows: Section 2 addresses the DT and its current research demand, behavior models, challenges that arise in this context during the machine life cycle and related work, finishing with a conclusion on these subsections. Section 3 describes the approaches of automatically creating DTs and adapting DTs, as well as bringing the concepts together over the machine life cycle. Then, in Section 4, the use of the behavior models and their benefits over the machine life cycle is illustrated using a modular production system. Finally, a conclusion and an outlook on future work are given in section 6.

## 2. Initial situation

Beginning with current research on the intelligent DT, the following section continues with a closer look on behavior models and its challenges in development and use. Following this, related work in this context is shown and a conclusion for the following section is drawn.

### 2.1. Intelligent Digital Twin

The DT is being used in a variety of fields, such as healthcare, smart cities or manufacturing [8,9]. The concept of DT, which dates back to its origins in the 2000s, has seen a significant increase in publications since 2016 [10]. Over the years, a variety of definitions for the concept have emerged [11]. In this work, the definition according to [12] is used extending basic data, models and relations by the capabilities of simulability, active data acquisition and synchronization. Fig. 1 presents the core components of such a DT in light blue. If the DT is extended by *Services*, *DT Model Comprehension*, *Intelligent Algorithms* and a *Feedback Interface,* all marked in dark blue, it becomes an intelligent Digital Twin

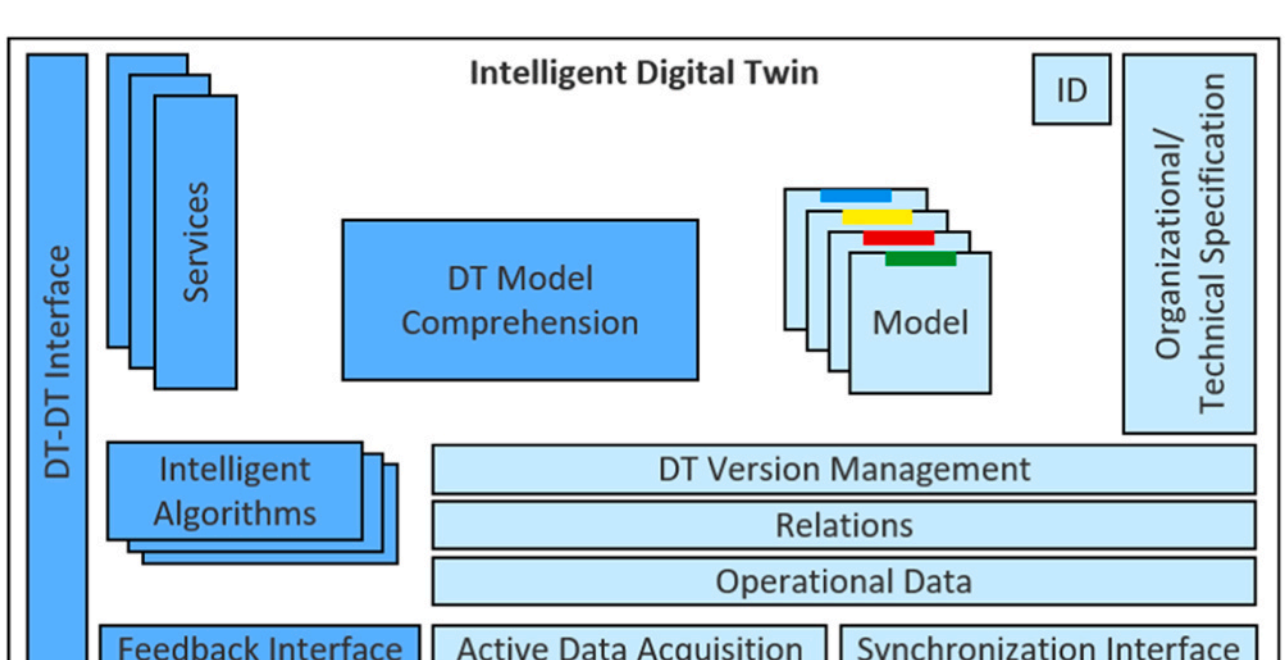

**Fig. 1.** Concept of the intelligent Digital Twin [12].

(iDT) [12].

The intelligence of the iDT can have many facets. For example, the automatic generation of control code for newly added machines within the framework of Plug & Produce, the optimization of the process sequence or predictive maintenance based on the *Operational Data* stored in the DT during runtime. [13].

The realization of such a DT or iDT faces many challenges, leading to research focusing on specific areas of the concept. In the following, an overview of some research areas and their challenges related to DT and iDT at the Institute of Industrial Automation and Software Engineering of the University of Stuttgart is given. The overview then serves to situate the research topic of this paper within the context of iDT.

In [14,15] an approach is presented creating a DT for existing production systems. He focuses on one basic principle of DTs, their model relations. The proposed automated approach analyses the information structure in heterogeneous sources (PLC code, process data, position data), combines them in a graph-based knowledge base and enriches it with semantics. Implicit knowledge is extracted with frequent subgraph mining [16].

Ashtari's [17] work focuses on the synchronization of DTs with the real manufacturing systems after commissioning, to keep the DT always up do date with the real world. Therefore, he proposed a method called Anchor-Point-Method, which uses anchor points in the mechanic, electrical and software domain. These anchor points are the component models with the smallest granularity of a mechatronic component or system in the three named domains. By keeping the DT always in sync with the real manufacturing system, it can be efficiently used for reconfigurations of the system. In an exemplary use case a time reduction of up to 58 % for the reconfiguration process can be achieved. [17,18].

In [19], the authors present a framework for intelligent industrial automation systems as a possible realization for the iDT. In the context of the iDT, a possible realization of the characteristics (observation, analysis, reasoning and execution) of intelligent systems is presented. A modular production system automatically reacting to new customer requests by AI-supported new control code based on the environmental parameters is used as a demonstrator.

Xia [20] focuses on methods for the automated data transformation based on neural networks to interpret the semantic meaning of textual data. Such data transformation methods are used to efficiently create the Asset Administration Shell (AAS) as a possible standardized data model for the DT from proprietary information models. For this purpose, he is currently working on an approach for the assisted generation of the AAS.

In [21] an industry-relevant use case for the DT is represented enabling the prediction of single tracks geometry produced by laser metal deposition. The DT provides data and models across companies. Both the machine manufacturers and the machine operators benefit from this bidirectional flow of information. The machine manufacturer gains access to operational data from real applications helping him to improve his behavior models. In return, those improved models are provided to the machine operators making their estimation more accurate.

In [22] it is shown that the DT needs up-to-date and high-quality data from the asset in order to simplify the increasingly complex functions and connections of industrial automation systems. The quality of the DT should be enhanced with additional data that exists outside the asset. To this end, a categorization of data for the DT and a methodology to identify and reuse the data are described.

The overview shows that there is a lot of specific research going on in the field of iDT. From the authors' point of view, we are still far away from exploiting the full potential of the intelligent digital twin, both in research and in industry. A future step will be to bring these individual research approaches, which represent only a small part of possible iDT functionalities, together in a suitable iDT environment.

As introduced in Section 1, another important area of research are behavior models in the context of the DT. The researches from [23] and [24] deals with the efficient creation, adaption and use of these models at certain points in the machine life cycle and improving their cost-benefit ratio. These behavioral models are briefly described below.

### 2.2. Behavior models

Behavior models are core components for a lot of different use cases of the DT such as virtual product development, virtual configurations, virtual commissioning or predictive maintenance [5]. Behavior models describe the dynamic, temporal and static behavior of systems, subsystems or system elements [25]. With quantitative mathematical models, the analysis of input to output behavior through simulations is possible [25].

However, a big challenge with behavior models is their variety, since different stakeholders from various machine life cycle phases partly understand quite different things when mentioning behavior models. In doing so, a behavior model can range from a simple input-output model, which merely depicts logical links of a software function, to complex 3D simulation models with several disciplines such as fluidics, mechanics or magnetics modeled using differential equations. This problem arises especially when using models consistently across organizational boundaries. In particular, knowledge about the different characteristics of behavior models is not universally available. Since this knowledge is one of the core aspects of DT, meta-information about the properties of the corresponding behavior models is needed. A structure of such meta information to enable an easy transfer of knowledge about the models characteristics is presented in [26]. This concept for structuring behavior models is shown in Fig. 2 and will be used in this contribution.

The categories of modeling range, modeling width, modeling depth, and the type of behavior are used to structure the properties of behavior models. In terms of modeling width, in many use cases there are the mechanical, electrical, and software disciplines that are common in industrial automation. However, there are also use cases in which other disciplines such as fluidics, magnetism or thermal behavior are relevant. These are grouped using the other disciplines block in Fig. 2 and can also be referred to as domain-dependent disciplines. The models of the individual disciplines can represent different scopes from the field device to the connected world. The depth of modeling is often directly related to the mapped scope. Comprehensive behavior models of entire stations, work centers or enterprises describe the behavior of individual functions rather abstract. The reasons for this are the lack of detailed knowledge, the lack of necessity of the corresponding detailing as well as the often negative influence of extensive detailed behavior models on the required computing power and the required simulation time. This connection arises from the different modeling of the behavior in the different modeling depths. Models of lower modeling depths (*discrete behavior*, *discrete temporal behavior*) often only model discrete signals between the inputs and outputs, sometimes with time delays. Such models can also be simulated efficiently with large scopes (high modeling range). The situation is different for behavior models with higher modeling depths (*physical non-spatial behavior*, *physical spatial behavior*). These model the behavior of the respective asset on the basis of differential equations, partly also of partial nature. If the scope of such models increases, the effort required to simulate such very extensive differential equation systems can quickly explode. Relatively independent of the modeling depth, the behavior of the assets under consideration can be modeled ideally or including error-prone behavior. By structuring the properties of behavior models in this way, the reuse of such models is possible in an efficient way even across organizational units and company boundaries since users can easily define what kind of behavior models they need in their use case.

In addition to the potentials of the behavior models, challenges also arise during their creation, adaption and use over the machine life cycle, mentioned in Section 2.3.

### 2.3. Challenges

In industrial automation, primarily machines and systems from the manufacturing industry are considered. A possibility of classification of use cases, requirements or similar can be realized by the time domain from development over utilization to recycling of the machine. The different phases during the life of such a machine are represented in the machine life cycle. The different phases in this vary depending on the point of view. For this paper, we will use a machine life cycle that is an overlap of [4,17,27–30], shown in Fig. 3.

The typical stakeholders participating in such a machine life cycle can be divided into three groups. Firstly, the actual manufacturers of the machine under consideration. These often manufacture the machine under consideration for a machine operator who uses the machine in his production to manufacture various products later in the machine life cycle. On the other hand, the machine manufacturer obtains a large proportion of the components, used in the machine, from component manufactures who have in-depth expertise in specific areas and produce a range of components for specific applications. The focus of most presentations on the machine life cycle is on the viewpoint of the machine manufacturer. This is also clearly shown by the phases in which the machine manufacturer is mainly active being the case for the first four phases of the machine life cycle until the machine goes into operation. From operation onwards, the machine operator of the machine is the main active stakeholder in the later machine life cycle phases. These different levels of activity over the machine life cycle for the different stakeholders are represented by the varying height of the bars in Fig. 3. During operation and maintenance, the machine manufacturer is mainly involved in repairs and the provision of spare parts.

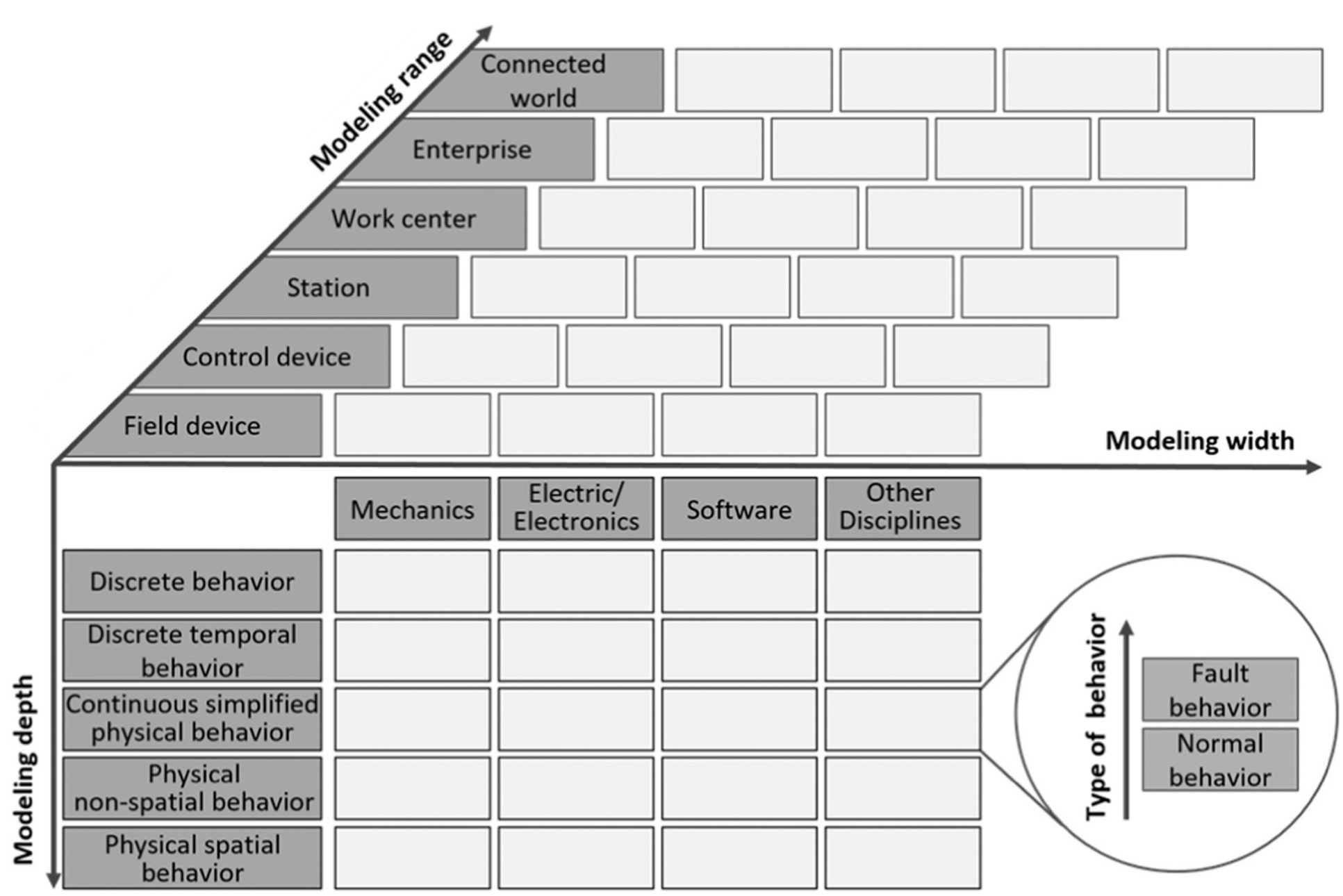

**Fig. 2.** Structure of metainformation for behavior models [26].

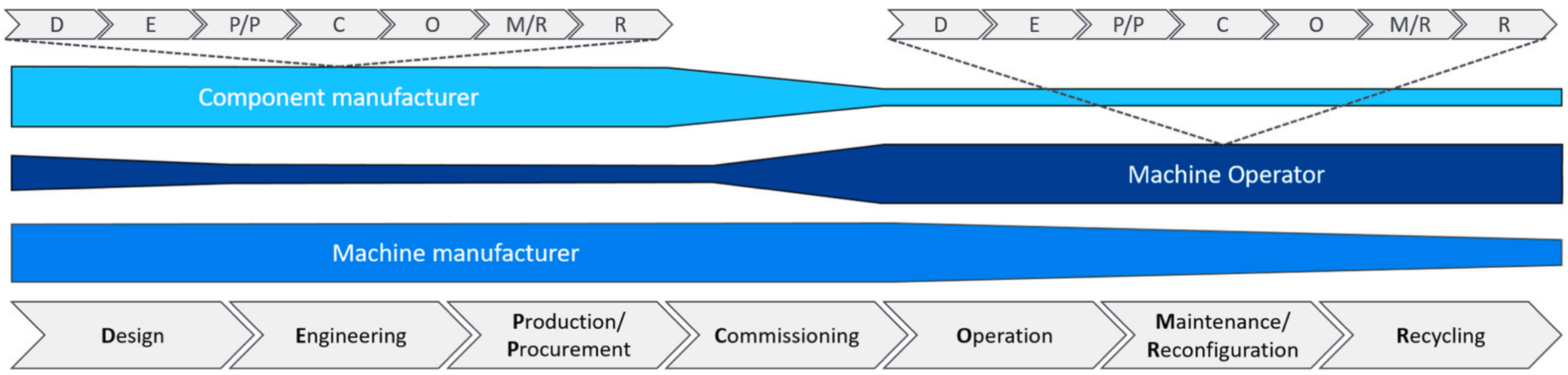

**Fig. 3.** Typical machine life cycle with active phases of typical participating stakeholder groups.

Parallel to these phases, the machine operator also has his own product life cycle. An example of this is the product development process of an automobile. In this case, the production machine, which is described by the machine life cycle shown here, only represent the phases from production preparation onwards. The early phases of concept and product development of the cars product creation process that are relevant before production is prepared are mostly detached from the machine life cycle of the production machine shown here. The sequence shown in gray above the operator illustrates this.

The same applies to the component manufacturer, who is mainly relevant in the design, development and operation phases of the machine life cycle. The components supplied for the actual machine have their own life cycle too, which overlaps with the machine life cycle only to a limited extent.

Looking at the different stakeholders along the machine life cycle, with regard to the efficient use of the DT, it can be stated that DTs need to be used across organizational units to avoid dedicated creation in each individual entity. With such end-to-end usage, the majority of the task of creating DTs is due to the component manufacturer [30]. The DTs supplied by the component manufacturer can then be aggregated at the machine manufacturer into DTs for the machine and passed on to the machine operator.

For all the benefits of the DT, its efficient creation is one of the key challenges [12,31]. If engineers create the DT manually, a large part of the achieved benefit is eaten up by its creation process [32]. In addition to the large amount of time required for manual creation, it is also very error-prone, which results in complex test procedures and additional time needed. Successful and value-adding use of the DT can therefore only be achieved through partially or fully automated creation [33]. Such automatic creation methods should be able to create models for internal as well as external use cases. [34,35]. These automated approaches must be easy to integrate into existing development processes [32]. On the one hand, this increases acceptance and, on the other, enables easy access to existing and created data and models from the existing development process. However, these existing data and models are often not available in a semantically unambiguous form [20]. Instead, mainly company-specific designations are used. For cross-company use, however, it is necessary to provide these data and models with semantically unique designations.

The various behavior models created by the component manufactures can be put into relation in the DT and used by the machine manufacturer, e.g. as part of a modular production system for virtual commissioning (VC). In this way, the Start of Production (SOP) date can be reached more quickly (Fig. 4), because errors can be identified at an early stage. [24].

As described above, the models are developed for specific applications and fulfill the respective requirements [36]. Therefore, from the authors' point of view, a direct coupling of models is not trivial. In order to ensure a profitable cost-benefit relationship, reality is abstracted for specific applications, so that the closeness to reality decreases and a so-called "reality gap" [37] results. Nevertheless, as described above, the models can be used for applications such as operation-parallel simulation, optimization or forecasting in the operational phase. In this phase, the mentioned "reality gap" increases due to the ageing process (Fig. 4), which is why the realism of the DT of an asset deteriorates if no model adaption is made. This ageing process is due to a variety of possible changes. On the one hand, new requirements can be placed on existing applications during the operational phase, which were still unknown in the development phase.

But new applications with their associated requirements can also be added. In addition, changes to the asset (e.g. unconsidered modification, replacement, wear) and in the environment of the asset (unconsidered influencing variables, e.g. temperature) can increase the "reality gap".

In order to counteract this ageing process, it is necessary to adapt the DT in the operational phase in order to maintain an application-oriented realism. The challenges of this model adaption are discussed below.

As already described, models of different components are increasingly being created for modular production systems, which in turn are provided by different component manufacturers [38]. These components include several mechatronic disciplines (mechanics, electronics, software) due to the increasing use of mechatronic components. A variety of simulation software has established itself on the market for the creation of models in the different domains [36]. But the type of modelling can also vary between white, grey and black-box approaches in order to answer the respective questions of an application [39]. Black-box approaches should not be confused with black-box models, which are used, for example, by component manufacturers during model transfer in order to protect their knowhow. For example, with the Functional Mock-up Interface (FMI).

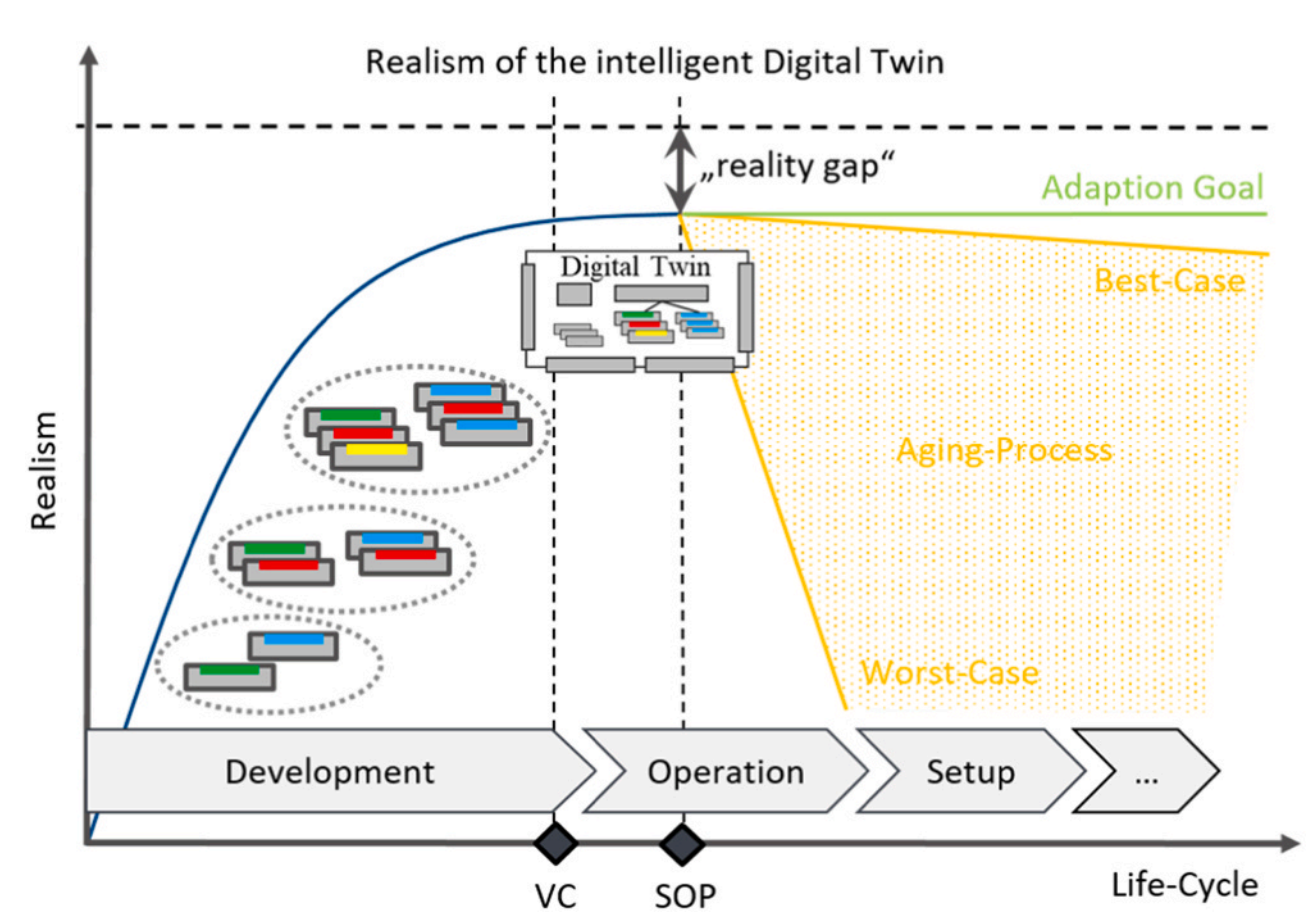

**Fig. 4.** Deviation of the realism of the DT over the life cycle.

Due to the heterogeneity described above, the authors assume that it will be very difficult to reuse the models in the operational phase. In the case of a necessary model adaption due to the aforementioned unforeseen changes, such as a change in the asset or from an application perspective, system knowledge is required. This means knowledge about dependencies and interactions within the heterogeneous model landscape as well as knowledge about the individual models in order to identify suitable models for the respective change. The model knowledge here usually lies at the creator of the model, such as the component manufacturer. In addition, there is the necessary interoperability to be able to exchange data between models as well as the necessary knowledge to maintain suitable model compatibility at system level [40]. It follows that the manual model adaption, especially for behavior models, is time-consuming, costly and error-prone. [24].

### 2.4. Related work

This section describes approaches from literature dealing with the creation of behavior models. Moreover, approaches in context of adapting and using models are listed.

#### 2.4.1. Creation of behavior models

There are existing approaches in literature for the low-effort creation of behavior models. In the following, the most relevant approaches from literature found by the authors are shortly explained.

Beginning with [41] an approach for the automated creation of models based on technical documents such as bus configurations or input-output-lists is presented. These documents are used in a method to create the basic functionality for behavior models of production machines.

[42] presents an approach in the field of virtual commissioning too. The MCAD2Sim approach is a structured-data-driven approach based mainly on 3D MCAD models. From these 3D MCAD models, different algorithms extract information about kinematics, components and geometry as well as orientation in 3D space.

In [43] the focus is on the machine manufacturers as in the two approaches above. In this approach, virtual building blocks are created from behavior building blocks and graphical building blocks, which are then coupled to form a virtual machine based on customer requirements. This virtual machine can then be used for hardware-in-the-loop simulations.

[44] present an approach for creating a DT for a factory. For this purpose, 3D scans and analyses are used, among other things, on an object recognition by means of Deep Learning. The models for the DT are created from the recognized objects with further information from external databases.

[45] present an approach for the automated creation of simulation models that are used for control code tests. The main focus is on given computer-aided design documents, especially piping and instrumentation diagrams. A similar approach is taken by [46], who also deals with virtual commissioning in continuous processes.

[47] proposes a process industry approach. Different sources of information such as preliminary process design, equipment data sheets, piping and instrumentation diagrams and 3D machine models are used to create a simulation model in an automated way. An online system architecture is proposed that enables faster system integration and the use of model optimization and online estimation methods.

[48] present an approach for the automated creation of models based on machine design. The knowledge for the models is stored in a simulation library, which can then be used by the creation algorithm. The focus is on modelling entire machines specifically for virtual commissioning.

In summary, however, most approaches in the literature focus on building behavior models with modeling depths of discrete behavior or discrete temporal behavior, mostly used in the context of virtual commissioning. No approaches for the creation of behavior models of modeling depth physical non-spatial behavior could be found in the literature by the authors. However, this modeling depth is mainly needed for applications in early machine life cycle phases.

#### 2.4.2. Use and adaption of behavior models

Various approaches exist in the literature dealing with behavior models, configuration and co-simulation. The most relevant approaches are outlined as follows:

The authors in [49] present an agent-based assistance system for automatically composing and configuring a co-simulation for the commissioning phase. In [50], the authors extend this with an operation-parallel model adaption. A modeled multi-objective optimization problem is solved using an evolutionary algorithm.

[51] describes the use of a multi-agent system (MAS) to implement an AAS. The authors in [52] illustrates a similar approach to orchestrate components such as models and services within the AAS as well as external communication via an agent-based interface.

[53] present a reconfiguration management architecture for generating alternative machine reconfigurations with agents. Simulation-based multi-objective optimization is used to determine optimum close reconfigurations, which are created using dynamic simulation model generation based on production requirements.

[54] describe a co-simulation framework in which the integration of heterogeneous models is addressed by coupling different simulation tools within a simulation environment at runtime. The agent-based approach allows the integration of non-standardized models and also the dynamics at runtime in the sense of "plug-and-simulate".

In [55], the authors present an approach that enhances the iDT with self-organized reconfiguration management. Nonlinear autoregressive exogenous models are extended by autoregressive modelling of anomalies in order to continuously improve the models with process data and thus enable to find the optimized reconfiguration alternatives more comprehensively.

A product life cycle management system with a process simulation for flexible assembly systems is presented in an integrated reconfiguration planning tool by [56]. An extended entity-relationship data model is used, which contains a product-process-resource, a simulation and a production programmed submodel. It supports the generation of the simulation model and the simulation execution for different planning alternatives.

In [57] the authors derive the requirements for semantic models to capture production simulation knowledge and present an ontological architecture for a knowledge graph for manufacturing and simulation. The knowledge base is the basis for the adaption as well as the automatic creation of simulations. This approach focuses on material flow simulation.

The authors in [58] use a formalized asset model for dynamic reconfiguration of simulations. They use semantic reflection to detect structural drift and monitor basic properties. The authors use Sematic Micro Object Language for their implementation of the DT infrastructure. The focus lies on the structural reconfiguration of DTs.

From the authors' point of view, no approach could be identified that deals with automatic model adaption over behavior models specifically in different modeling depths of modular production systems. However, the approaches show that agents and ontologies are promising technologies for the realization.

### 2.5. Conclusions from the initial situation

Based on the previous sections, the following conclusions can be drawn for the remaining sections of this paper:

- The concept of the iDT as well as the intelligence within it has many facets and represents a need for research.
- Behavior models are one of the central components of the DT and can provide benefits in many use cases across the entire machine life cycle.
- There are a number of different behavior models that vary widely in terms of calculation, creation, adaption, and usage effort.
- Low-effort creation is one of the key challenges of the DT being particularly important for component manufacturers for consistency reasons.
- Approaches for creating DTs exist, but not with a focus on different modeling depths for use cases across machine life cycle.
- The adaption and use of the models as well as their coupling in the context of modular production systems requires expert knowledge and is very time-consuming.
- In the context of model adaption, no approaches were identified for different modelling depths and different applications, but agent and ontology approaches are promising for implementation.

In the following, concepts from the authors' work are presented and brought together for the first time to address the aforementioned challenges. Here, the focus is on the interfaces between the two concepts. This is intended to make a contribution to the use of behavior models in the DT over the entire life cycle. To illustrate the possible added value through the consistent use of behavior models from DT along the entire machine life cycle, industry-related use cases with corresponding results are presented below.

## 3. Approaches

In this section, the approaches for low-effort creation of DTs (Section 3.1) and for an automatic adaption of DTs (Section 3.2) are presented. Section 3.3 unites the concepts and places them in the context of the iDT described to enable efficient use of DTs over the entire machine life cycle.

### 3.1. Low-effort creation of Digital Twins

Since the efficient creation of DTs is particularly important for component manufacturers, this section focuses on them creating DTs. The general concept to efficiently create a DT structure is shown in Fig. 5.

Requirements for creating a DT can come from external or internal sources. These heterogeneous requirements are preprocessed and harmonized in a first step shown in the *requirements preprocessing* step in Fig. 5. For the requested DT, on the one hand, the existing data and models from internal company sources must be attracted and semantically processed. Since the product portfolio of a component manufacturer often consists of a limited number of variants, an initially created translation table can be efficiently used for this purpose [33]. This translation basis contains mappings between enterprise-specific designations and their semantically unique counterpart from accepted libraries such as ECLASS. If, on the other hand, a more flexible approach is needed, methods from [20] can be used. This process of attracting and semantically preparing existing data and models is shown on the bottom part of Fig. 5. The primary information needed for the semantic preparation of existing data comes directly from company-specific IT systems such as product-lifecycle-systems (PLM), product-information-management-systems (PIM) or other systems already involved in existing development processes.

This semantic processing of already existing data and models should not be the focus of this concept. Rather, the focus is on the creation of currently often non-existent behavior models in relevant modeling depths. This process is shown in the upper part of Fig. 5. Similar to the semantic preparation of existing data, the behavior models are created using a library. The models out of this library need to be parameterized, therefore a data basis is used. Both, the data basis and the library of the behavior models are created in advanced. Both are domain specific and contain a lot of existing knowledge from the component manufacturer. While the information for the data basis can be prepared relatively easy in an automated process using measurement data, catalog data and many more, it is different for the behavior model's library. The behavior models of the corresponding components from the product portfolio are available in the library at field device level so that the method shown in Fig. 5 can create DTs of field devices or combine them to DTs of systems on the level of control devices. However, the large number of variants in a component manufacturer's product portfolio is often problematic. If the models for all components had to be created manually, this would mean an enormous effort. However, a fully automated method for the creation of the behavior model library often fails due to the lack of primary information about the behavior [32]. In order to solve this challenge, a more in-depth view at the product portfolio of a typical component manufacturer helps. This shows, that the combination of components from the product portfolio to systems results in a huge variant diversity. The same applies looking into the opposite direction. A large number of components are based on a significantly smaller number of basic building blocks and an even significantly smaller number of basic physical principles [33]. The illustration of this relationship is shown in Fig. 6.

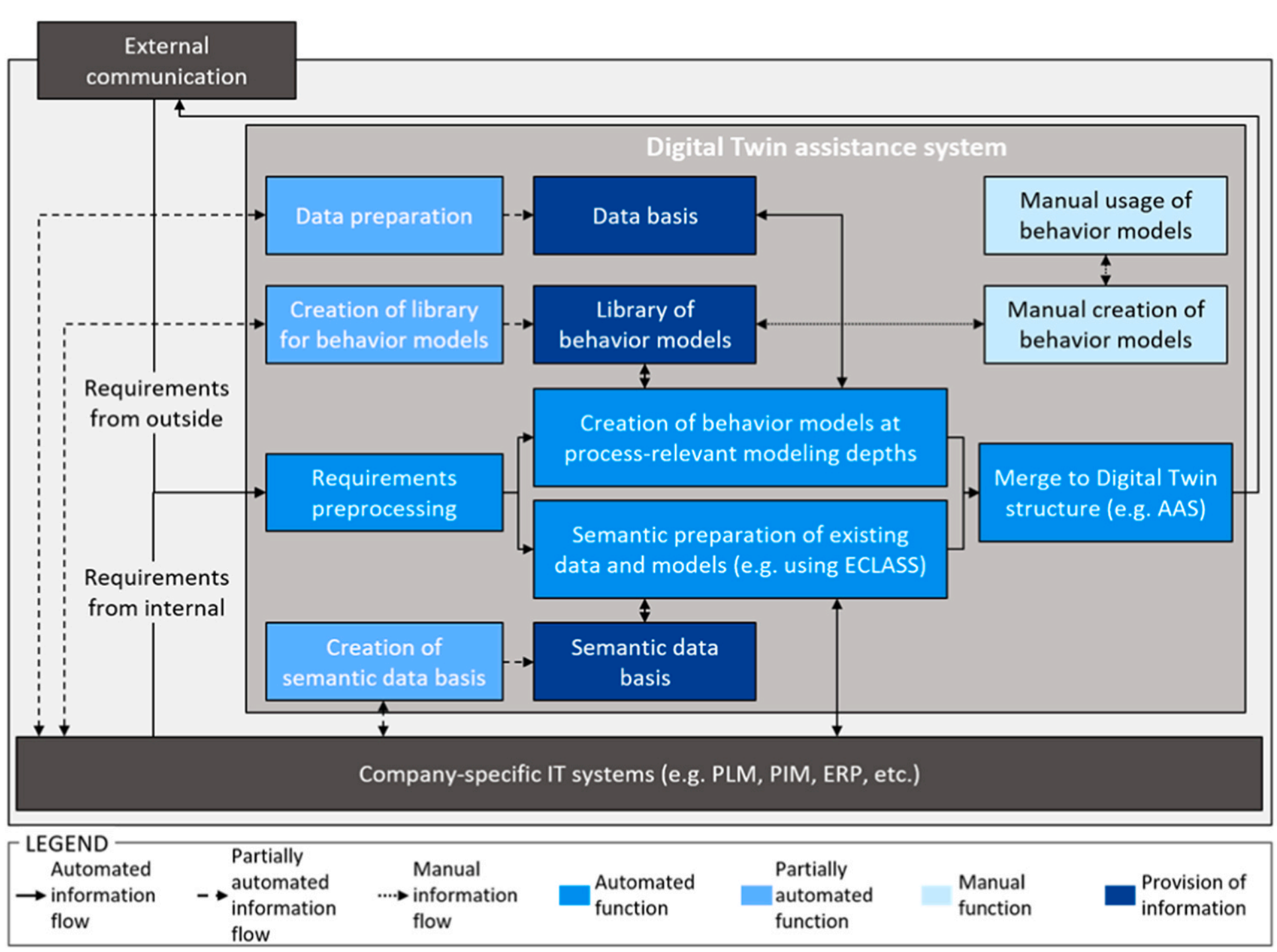

**Fig. 5.** General concept to efficiently create a Digital Twin structure.

The modeling range from Fig. 2 is shown on the left of Fig. 6. Compared to this, the blocks in the center show a typical structure for the product portfolio of a component manufacturers. Exemplary representatives of the respective groups are shown for each group. Central in Fig. 6 is the visualization of the relationship that the variance basically increases from the basic physical principle towards the system. However, it is important for a low-effort creation of behavior models for DT to keep the effort for the creation of the behavior model library from Fig. 5 as low as possible. The goal must be manual effort, if possible only for the production of *basic physical principles* or *building blocks*. The behavior model library for the use in the concept on level of the components can be provided again by the use of parts lists, circuit diagrams or other information sources (partly) automated. For this purpose, the concept from Fig. 7 can be reused, explained in the following. With the help of the created behavior model library, an efficient creation of process-relevant behavior models is possible. After the fully automated creation of behavior models and the semantic processing of existing data and models, this information can be combined into a DT structure, as shown in Fig. 5 on the right. Since DTs are primarily intended to be used universally and thus across companies, the use of standards for the DT information structure is advantageous. The AAS is used as a standard in this concept as an example. The DTs created in this way can be used for use cases in other companies or for internal company use.

This flow is optimized for use in defined processes with a known system structure. However, since there are also frequent use cases in the development process in which, for example, different system structures are to be compared with each other or test setups are to be mapped digitally, the assistance system also offers the option of creating and using the behavior models manually. This is intended to make the models and data of the behavior models easily and flexibly accessible to non-simulation experts. Since the automated creation of the behavior models is the central part of the concept, it will be explained in more detail in Fig. 7. It involves a more detailed view into the box to create the behavior models at *process-relevant modelling depths*.

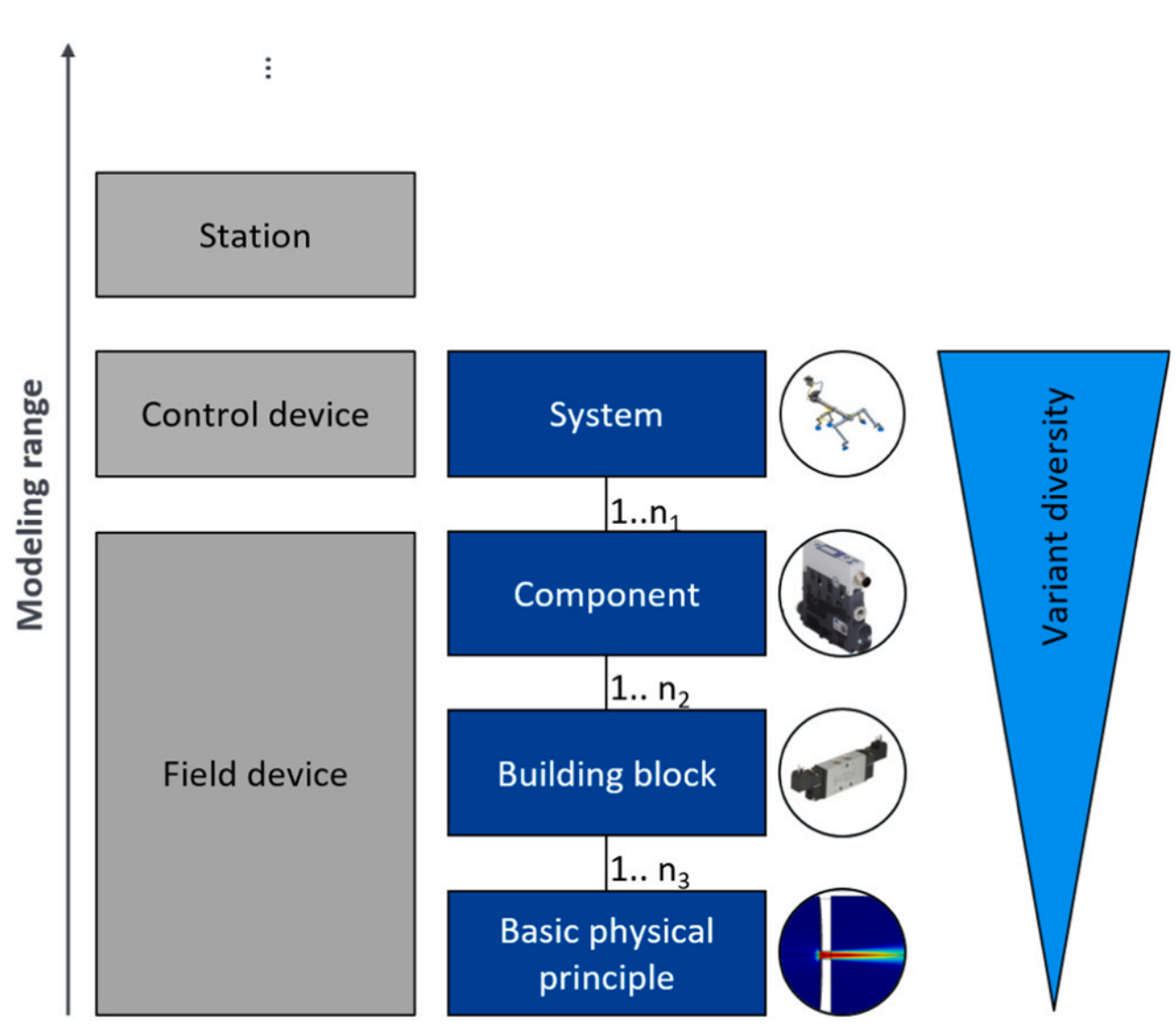

**Fig. 6.** Substructure of modeling range used to efficiently create behavior model library.

The requirements for the creation of a behavior model in a process-relevant modeling depth are fed into the method shown in Fig. 7. First, the information supplied in the model are analyzed for semantic uniqueness. A possible structure for the information is a graph [33]. If the information is semantically unique, they can be used directly. If this is not the case, the information must first be interpreted with a company-specific database. Then, using the type ID from the respective nodes, the corresponding behavior model modules can be attracted from the library. These modules can then be linked using the information from the edges. If the goal is to create a type-specific model, the parameters are attracted from the database using the type ID, in the case of an instance-specific model by the instance ID. In the last step, the model created and parameterized in this way is converted into a standardized format so it can be used in different simulation environments. FMI will be used as an example in this article. The behavior models created in this way can then be transferred to the AAS. In this context, the AAS can serve as a standardized interface for the transfer of the behavior models to other stakeholders who are active in further machine life cycle phases, such as the operational phase.

### 3.2. Model adaption in the intelligent Digital Twin

As described in the challenges (Section 2.3), machine operators face the problem to adapt and use behavior models in the operational phase. A need for continuous automated model adaption in the operating phase is shown, in order to use the potential of the models created in the development phase with minimal manual, time-saving and error-free effort. A novel concept for the model adaption is presented below. For a more detailed description and a first realization of the concept with a MAS, the authors refer to [59]. Here, a summary overview will be given in order to focus on the combination of the two concepts over the machine life cycle (Section 3.3). Fig. 8 gives a general overview of the boundary conditions for model adaption (blue) and the required functionalities (gray).

The individual blocks and their interrelationships are described in the following. As already mentioned in Section 2.3, a change leading to a reality gap (drifting apart of virtual and real behavior) can occur in the operational phase through several scenarios. On the one hand, a change can occur through the addition of new *Applications* (e.g. optimization, prognosis) with new requirements, but also through the change of requirements (e.g. real-time capability) of an existing application in terms of *Time, Quality or Costs*. In addition, a change can occur in the *Asset* (e.g. *Abrasion*) or it's *Environment* (e.g. temperature).

In order for the *Model Adaption* to respond to these scenarios, it first needs information from the *Comparison of real and virtual behavior*, which compares the data from *Active Data Acquisition* for the real behavior with the virtual behavior of the applications. In addition, the model adaption needs meta-information about the *Active Model Configurations* to enable a comparison with the requirements of the applications. In order to efficiently find a suitable model configuration for the above-mentioned scenarios in a heterogeneous model landscape, the concept is based on the Plan-Do-Check-Act method (PDCA). This is an established method from lean management for continuous improvement and enables a systematic approach to solving complex tasks. Fig. 9 shows the method in more detail. The individual steps are briefly explained below.

In the *Plan* step, the need for adaption is determined based on the comparison of requirements and the comparison of real and virtual behavior leading to the definition of an adaption goal. In the *Do* step, a two-step adaption approach (structural and parameter adaption) is used to generate possible model configurations from a heterogeneous *Model Pool*. The structural adaption refers to the orchestration of different models from the Model Pool, e.g. in different modelling depths.

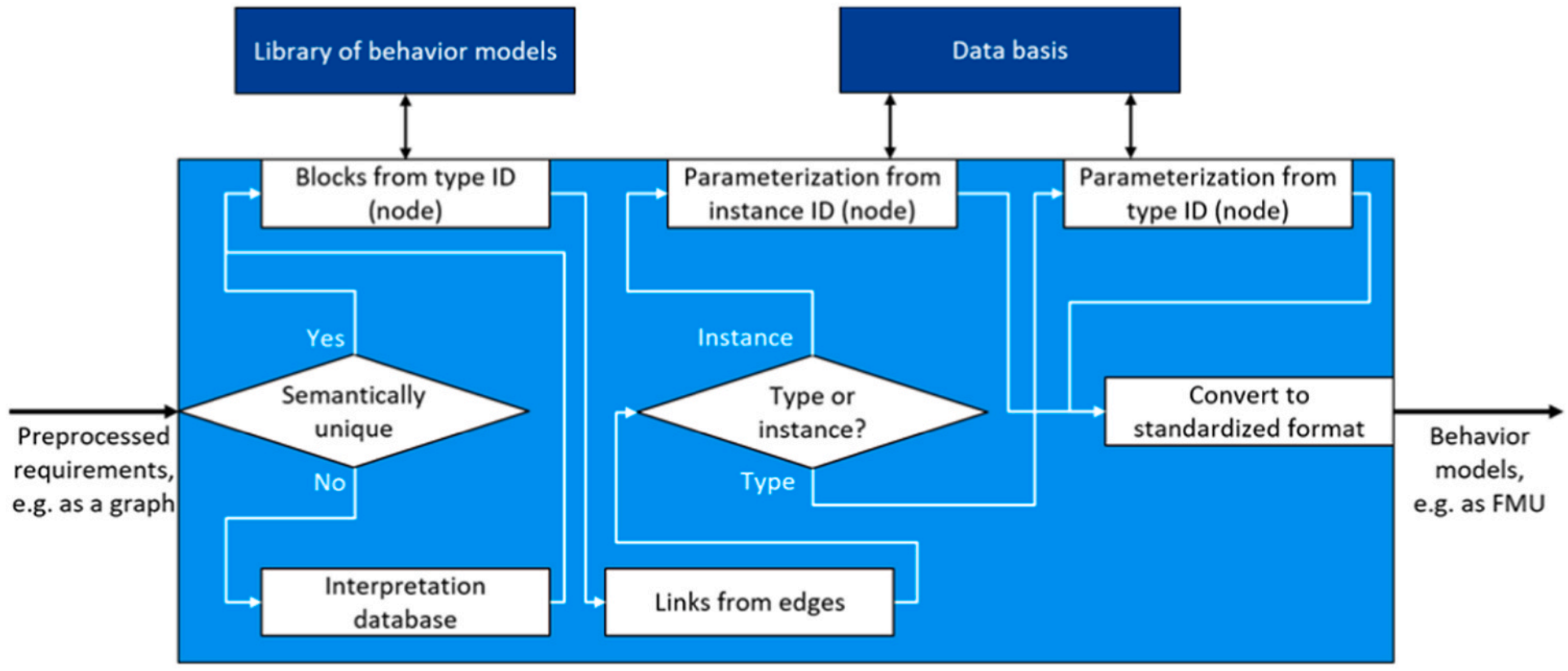

**Fig. 7.** Concept for automated behavior model creation.

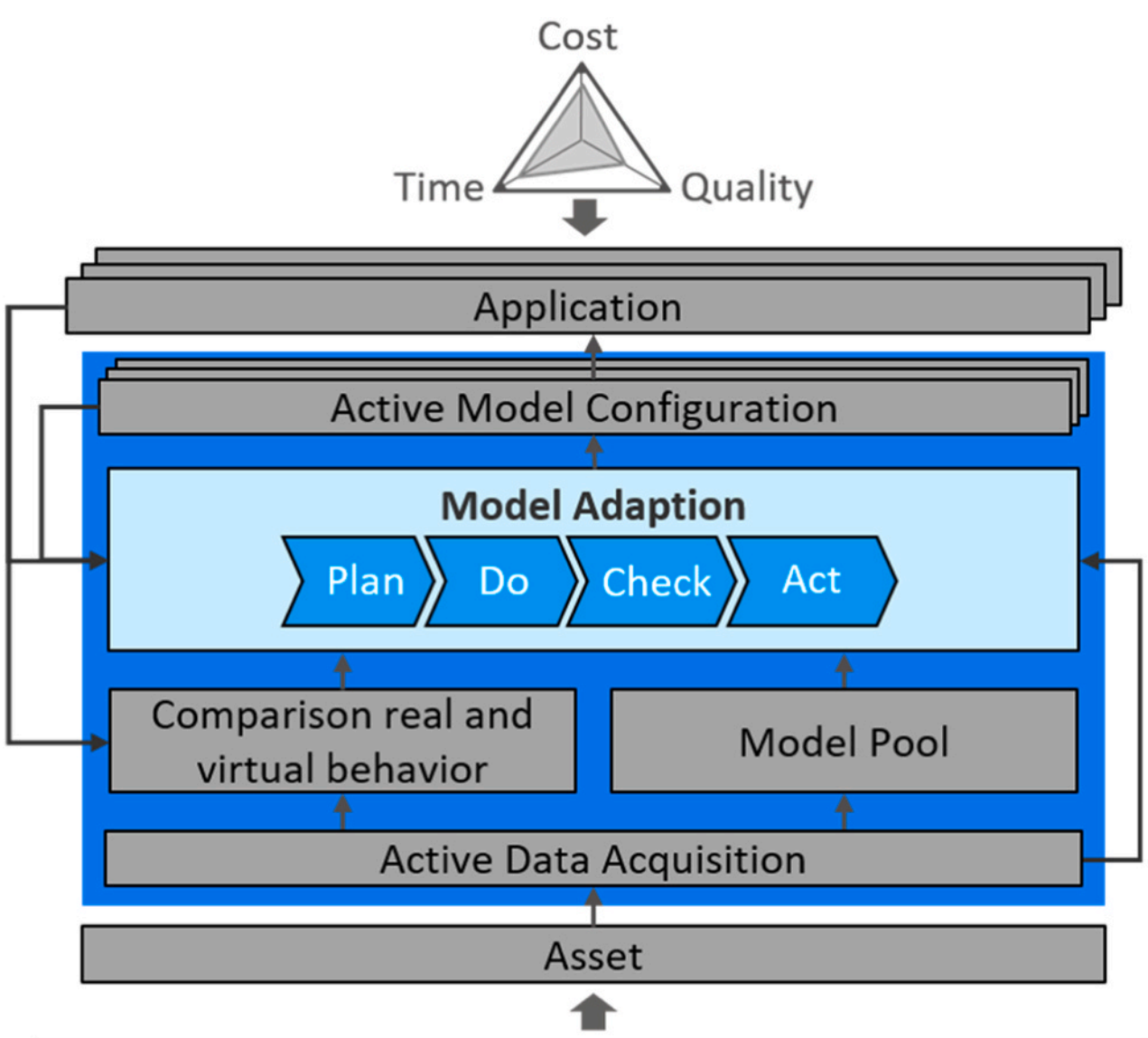

**Fig. 8.** Overview of the model adaption approach.

The generated model configurations are transferred to the *Check* step and evaluated with a utility analysis with regard to time, costs and quality. For the evaluation of the quality, the comparison with the real data from the *Active Data Acquisition* must take place. The most suitable model configuration for the specific application is activated in the *Act* step. The use of the PDCA cycle makes it possible to structure the complex solution space that arises from a heterogeneous model landscape and to efficiently determine suitable models for the respective scenario. However, manual execution can be time-consuming and error-prone, which is why model adaption should be done automatically. In [59] a possible realization with agents is presented. The concept is realized with JADE, which is a suitable framework for a MAS. The logical steps of the PDCA cycle are currently realized by individual agents. This enables the individual steps to interact with the upstream and downstream steps in a goal-oriented and proactive manner. Individual models in different modeling depths are encapsulated with so-called partial model agents.

These partial model agents have meta-knowledge from the proposed model structure in [26] necessary for model adaption. Due to the persistent and proactive behavior of the agents, the communication effort can be reduced. The individual PDCA steps only become active when required. Furthermore, only the partial model agents of the models that fulfil the defined adaption goal of the *Plan* step create model configurations in the *Do* step.

### 3.3. Relationship of the concepts

To bring the two concepts together the machine life cycle can be used, shown in Fig. 10.

In the early phases of the machine life cycle, DTs of individual components and subsystems are created at the component manufacturer and used for designs, validations or similar purposes and enriched with data. During the manufacture of these components and subsystems, further information is generated, which are also stored in the DT. In the process, the DT is mainly enriched in the model pool as well as in the technical data and with initial operating data. After manufacturing at the component manufacturer, this information can be passed on to the machine manufacturer, who uses them in its machine building process.

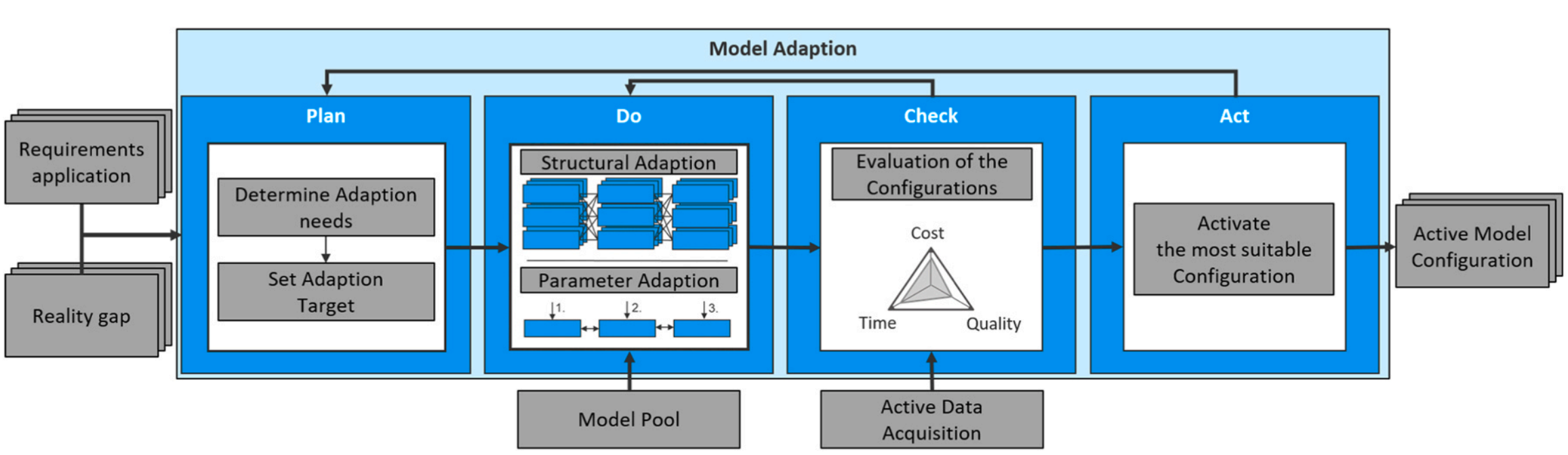

**Fig. 9.** Model Adaption Concept based on Plan-Do-Check-Act Cycle.

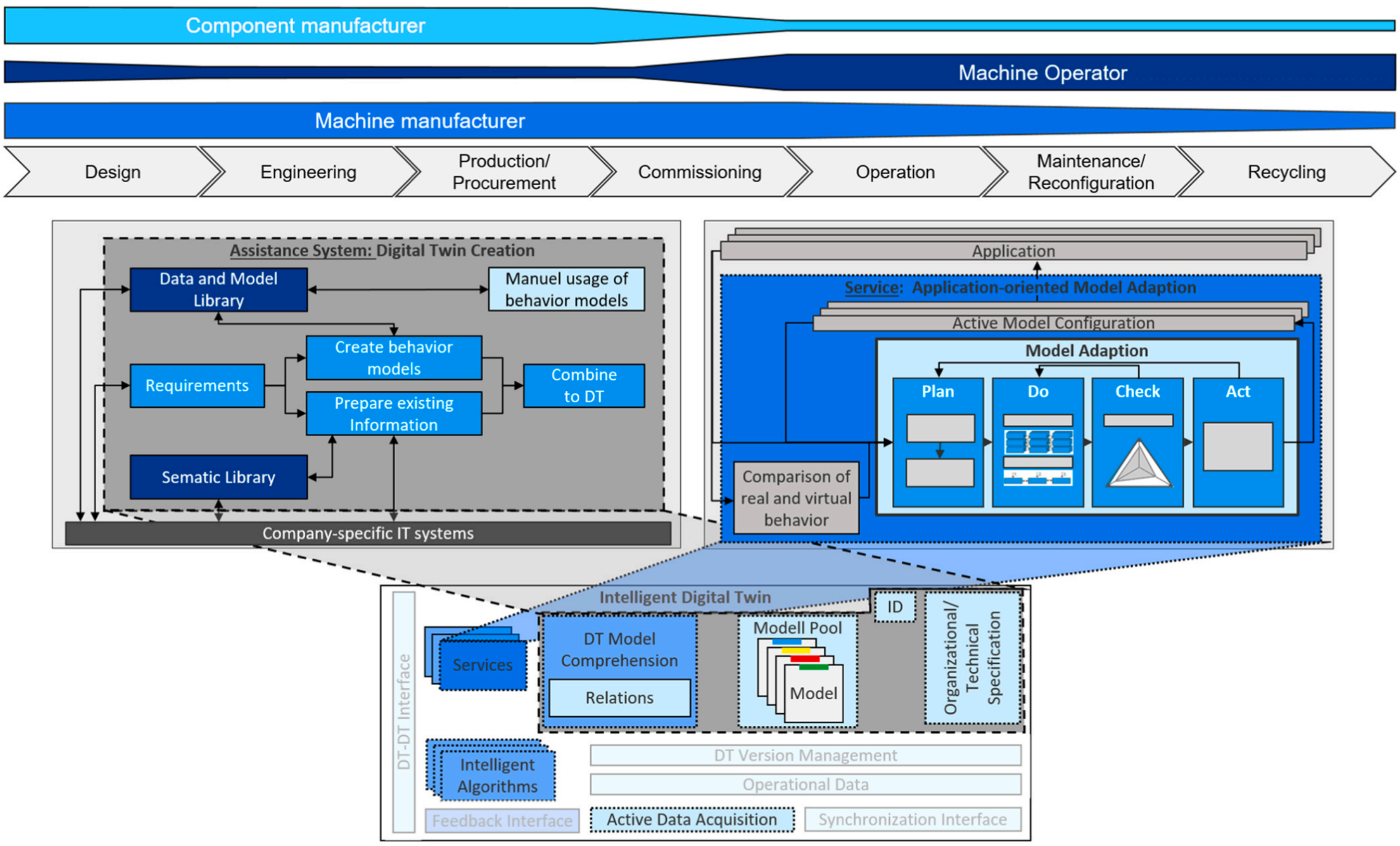

**Fig. 10.** Combination of efficient creation and usage of iDT over life cycle.

After completion, these machines are put into operation and handed over to the machine operator. In both transitions between companies, a simple and efficient exchange of information is possible via the DT. Especially when the exchange of information is compared with alternative options such as data sheets, Microsoft Excel files or via a homepage. The DT is generated and provided by the component manufacturer. If this DT is extended by aspects such as services or the DT Model Comprehension, the DT becomes a iDT, as described in Section 2.1. In this context, only the boxes of the iDT from Fig. 10 that are not grayed are relevant for the concepts described in this article. The other boxes are necessary for the basic functionality of the iDT, but not core aspects of the concepts presented here.

As soon as the equipment goes into operation, the model adaption becomes active. Based on the information from the Plan step, the Do step requests models from the DT M*odel Pool*. In addition, the *DT Model Comprehension* provides the necessary knowledge about the models and possible model configurations to the Do step. The DT Model Comprehension contains, for example, information that can be derived from the structuring according to [26] (e.g. model depth, model scope, model width).

In addition, information about model accuracy, model validity and model runtime are relevant to find a solution according to the 'magic triangle' described in Section 3.2. This requires a *Comparison of real and virtual behavior*. The real behavior is captured by the *Operational Data* of the *Active Data Acquisition*. The output is an *Active Model Configuration* that stores the most appropriate combination of sub-models for the specific *Application*. Intelligent algorithms (e.g., a genetic algorithm for improved model configuration identification in the Do step) can be used to support the model adaption, e.g., to implement learning capabilities from previous adaptations or to make the individual PDCA steps more intelligent.

The coupling of the concepts presented here thus enables efficient provision and use of the DT along the entire machine life cycle, even across company boundaries. The use cases that arise over this time horizon are very diverse. An overview of possible use cases from literature and their basic chronological classification into the machine life cycle is shown in Fig. 11 [2,4,5,7,29,34,44,47,60].

However, with the variance of these use cases addressed, two trends can be identified. First, the amount of data and models is increasing over the machine life cycle. If only models and basic information such as performance data are often known early on at the component manufacturer, this scope increases sharply when the machine enters the operational phase. The collection and provision of data and models is one of the central potential benefits of the DT too, since component manufacturers currently have only limited access to data from their components in operation, for example. Second, the required modeling depth of behavior models varies substantially over the machine life cycle.

While very detailed and extensive models are often required in the engineering phase, this changes as soon as the components are commissioned or enter operation. Then, fast executable models are preferably used, representing a less detailed behavior. A similar decrease in modeling depth can also be seen in the very early phases, when mainly basic designs and concepts are created.

The central aspect of this paper is not only the combination of the two concepts presented, which simplifies the consistent use of the behavior models from the DT along the entire machine lifecycle and across company boundaries. This paper also aims to illustrate the benefits of behavior models from the DT using industry-relevant use cases along the entire machine lifecycle. This illustration will take place in the following chapter.

## 4. Realization and validation based on a test set-up and an industrial application

The authors illustrate and validate the proposed concepts using the example of a robot arm equipped with a vacuum gripping system in a modular cyber-physical production system. For this purpose, the authors use behavior models in different modeling depths, modeling widths with different involved disciplines and errors in. One use case is presented for every machine life cycle phase. This large number of use cases is intended to illustrate the versatile use of behavior models with industry-relevant examples. It also shows why it is necessary to classify behavior models on the basis of the structure presented. The behavior models can be created and adapted individually for the respective use cases. However, this would be very time-consuming and inefficient. By using the concepts presented, the behavior models can be created efficiently for DT and then provided consistently over the machine life cycle for the respective use cases. Fig. 13 illustrates the use cases in each life cycle phases and their boundary conditions. The different properties are classified using the structuring from [26].

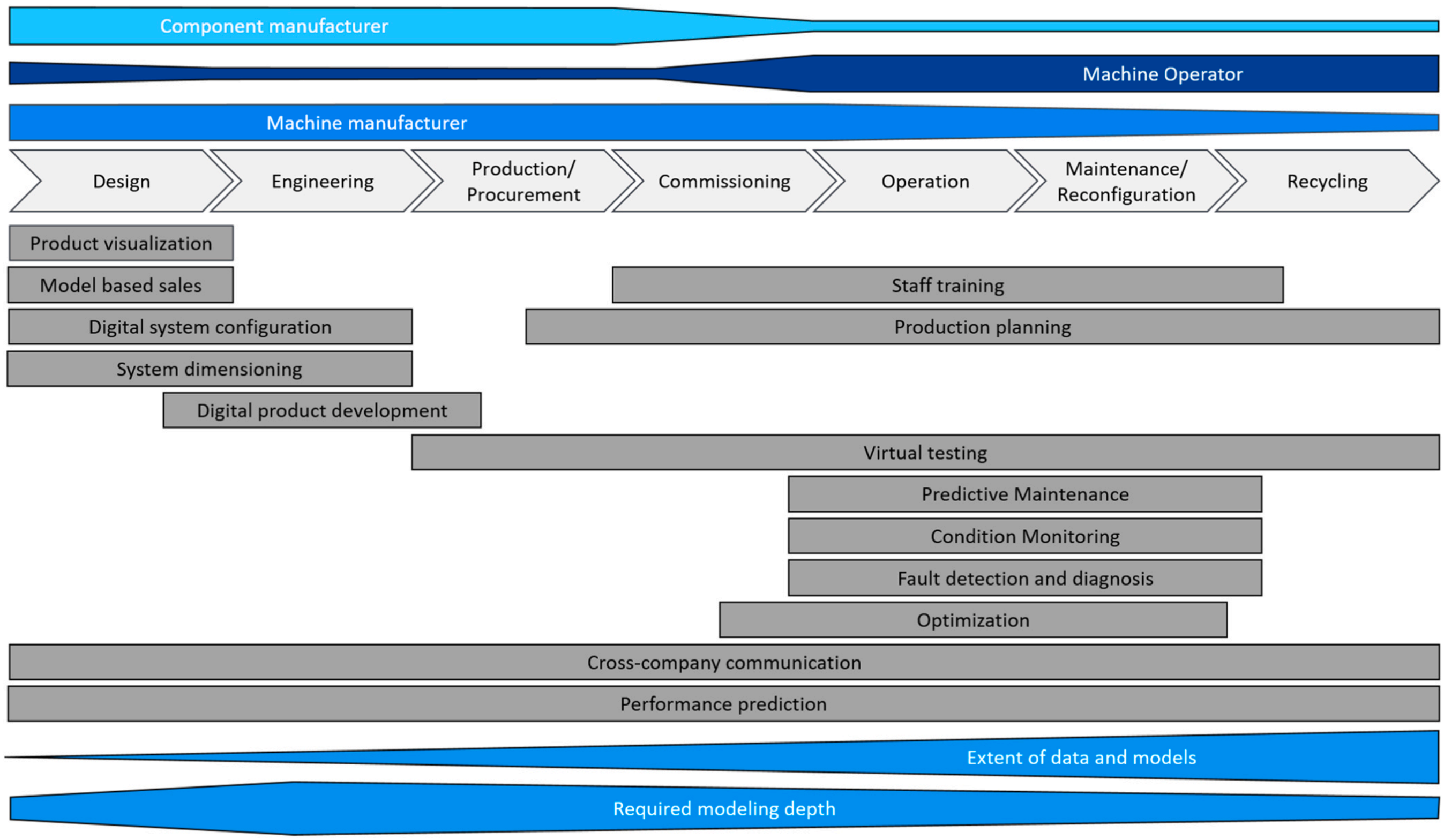

**Fig. 11.** Use cases over life cycle phases.

### 4.1. Demonstrator used for the evaluation

The authors use a modular cyber-physical production system, shown on the left side of Fig. 12, at the Institute of Industrial Automation and Software Engineering at the University of Stuttgart to describe the use cases over the machine life cycle. The system consists of a warehouse, five machining modules with integrated machining stations and four automated guided vehicles (AGVs). The AGVs take over the order-specific transport of products to the respective processing modules. The processing modules are equipped with conveyor belts to transport the products to the order-specific processing stations. One of these processing stations is equipped with a KUKA youBot, which uses a modular vacuum gripping system from J. Schmalz GmbH shown on the right side of Fig. 12.

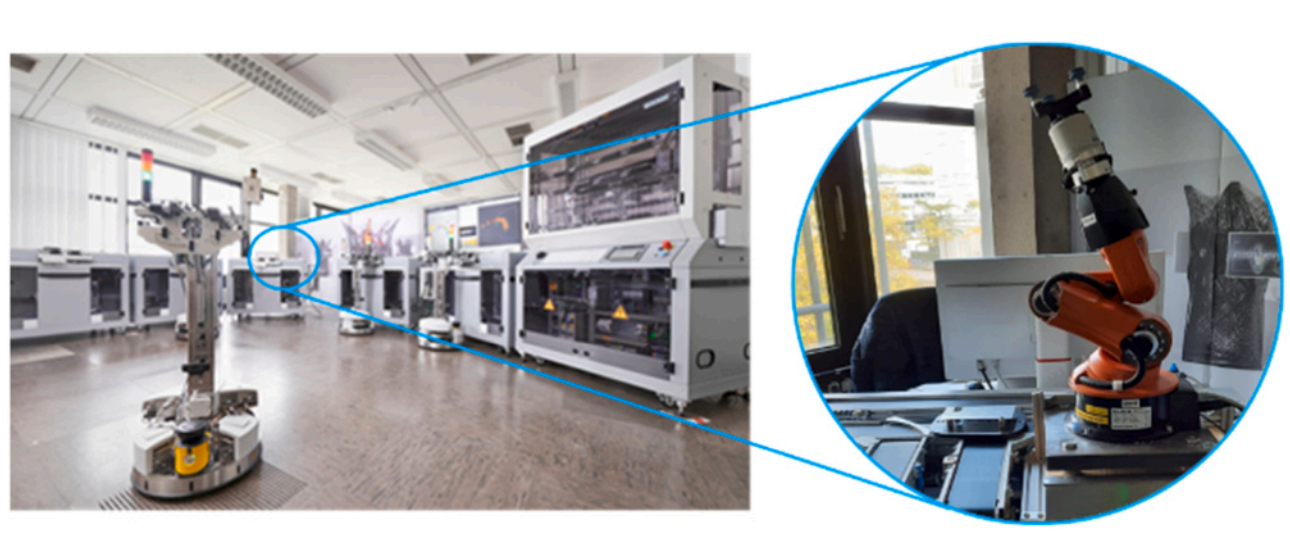

**Fig. 12.** Modular cyber-physical production system (left) containing the robot arm with a vacuum gripping system (right).

In the sense of this DT, behavior models in different modeling ranges, widths and depths of the robot (e.g. using GAZEBO and Simscape) and gripping system (e.g. using MATLAB, Simulink and Simscape) were created for different use cases.

The Robot Operating System (ROS) framework is used to control the robot gripping system. ROS controls the physical simulation of the robot in GAZEBO and the robot in parallel, thus enabling Synchronization. The Active Data Acquisition, e.g. of additional information such as the vacuum of the gripping system is realized using IO-Link.

### 4.2. Heterogeneous model landscape

An example for the models used in the various use cases is illustrated in Fig. 14 [26]. Therein, the models of the used vacuum pump with hoses and suction cups are shown for different exemplary use cases in different modeling depths. *Discrete behavior* models describes the vacuum gripping only with set and reset blocks concerning the digital input and output signals.

*Discrete temporal behavior* models models extend this model through delay blocks enabling a temporal behavior modeling. Changing to *continuous simplified physical behavior* models, the solely discrete signals from the *discrete behavior* or *discrete temporal behavior* models are extended by simplified continuous curves coming from simplified mathematical descriptions such as rules of thumb. A larger gap exists at the transition to *physical non-spatial behavior*. Here, not only input and output signals are modeled, but all components are modeled by physical equations. However, these are not spatially extended. This is supplemented in *physical spatial behavior*. Except for the modelling depth *physical spatial behavior*, all other modeling depths describe the behavior of the entire system. Such a complete system modeling is necessary for our use case too. Therefore, the authors consider the modeling depths *discrete behavior* to *physical non-spatial behavior*.

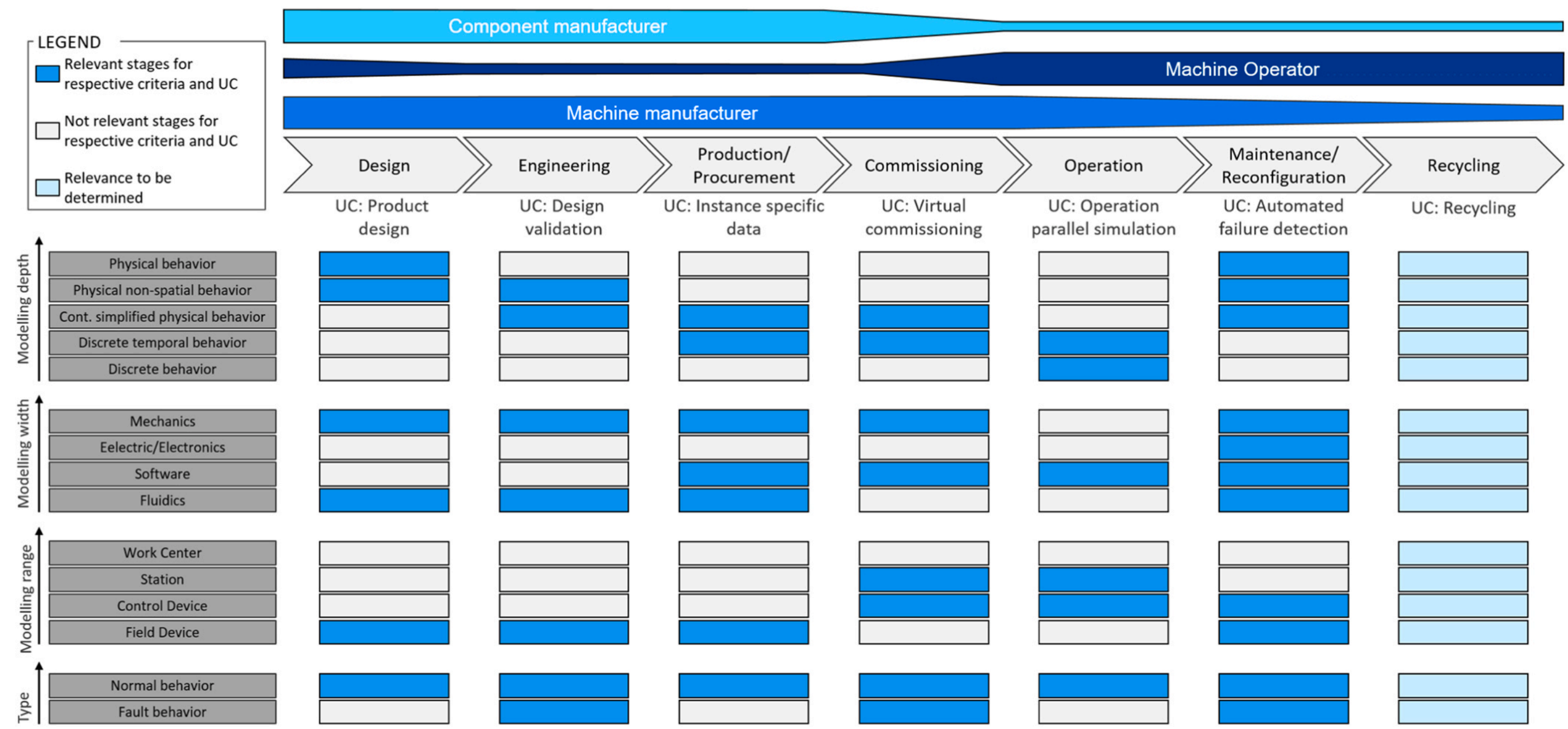

**Fig. 13.** Use-Cases over life cycle with corresponding behavior model properties.

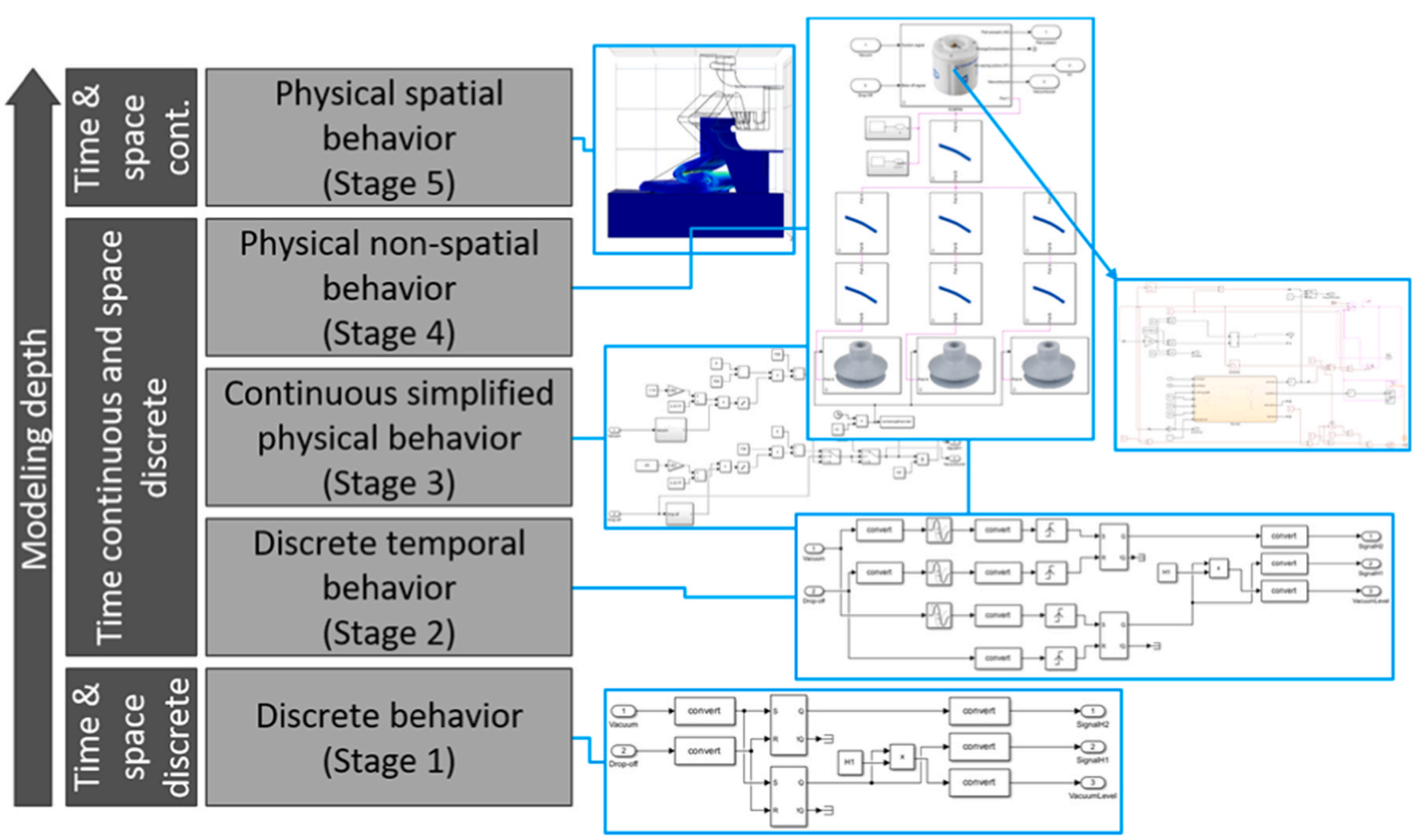

**Fig. 14.** Models used in the different use cases.

In general, a vacuum gripping system can be represented in a simplified way with its two inputs *Suction signal* and *Blow-off signal* and two outputs *Part present signal* and *In control range signal* shown in Fig. 15.

Even with just this one component of the overall system under consideration, there is a large number of influencing parameters. These can be divided into system parameters and external parameters. More and more of these parameters can be considered by using models of a higher modeling depth. Non-cyclic outputs can be used to monitor the vacuum level or the current power for example. These parameters can be simulated by the behavior models too.

The component models mentioned above, can be described with regard to their modeling depth as follows. Modeling depths 2 and 3 of the component models were realized in separate Simulink models. The duration between activating the *Suction signal* and a response at the *Part present signal (H2)* for modeling depth 2 is modeled using a time delay block. The parameters of modeling depth 3 are realized using an approximated equation with mathematical building blocks from [61]. Thereby, modeling depths 2 and 3 are modeled by unidirectional information flows. In contrast, modeling depth 4 is based on bidirectional information flows using the extension toolbox Simscape.

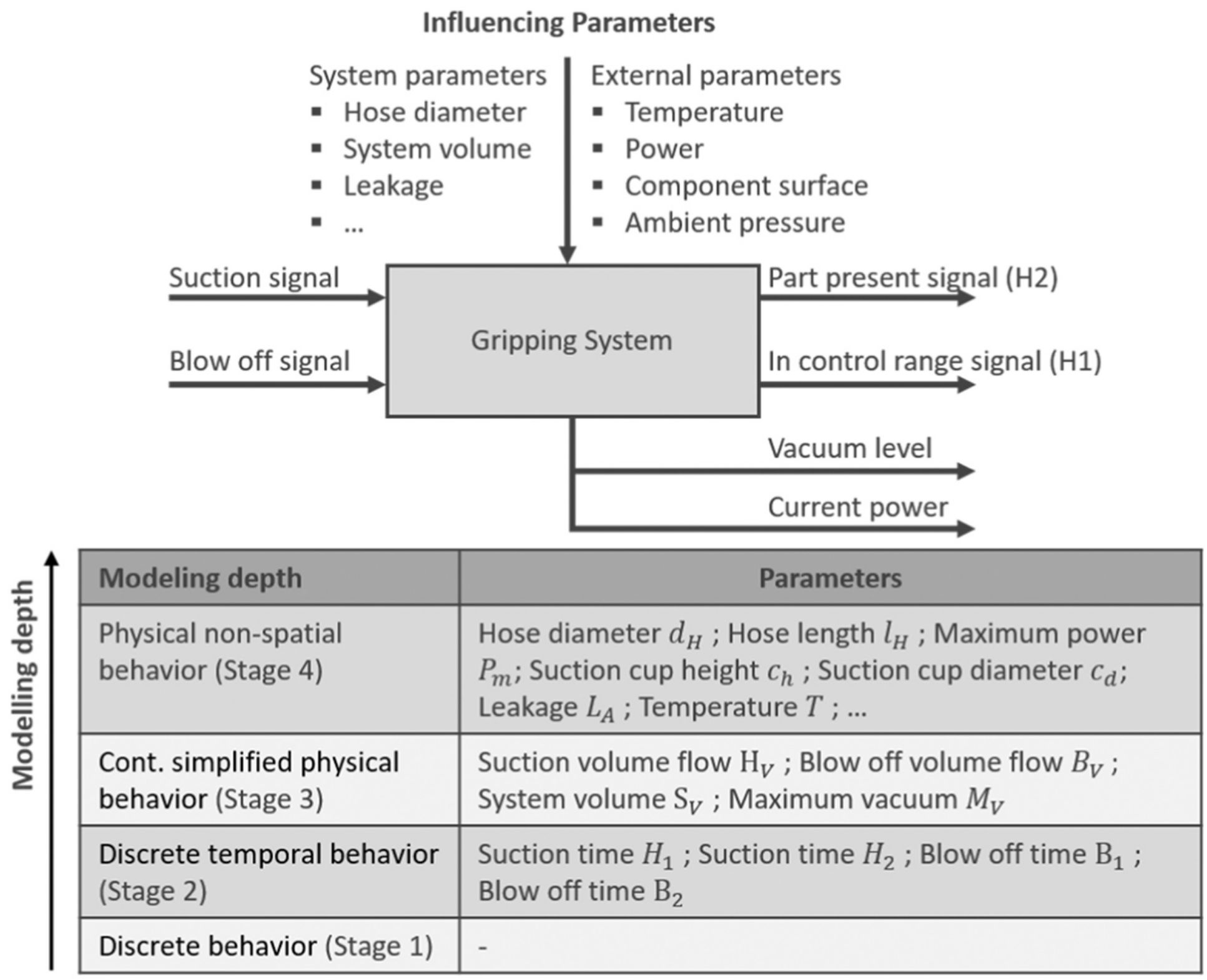

**Fig. 15.** Inputs, outputs and influencing parameters for the vacuum gripping system.

The components of the gripping system (pump, hoses and suction cups) and their physical behavior were modeled using differential equations. In addition to the parameters for the models shown in Fig. 15, there are parameters needed by the real system too. In this application, these are the switching thresholds H1 and H2 as well as the hysteresis h1 required for H1. An explanation of these parameters and the input and output signals is shown in Fig. 16.

It shows the temporal course of the input and output variables of the gripping system as well as some internal values. As soon as the suction signal starts, as shown in the upper plot, the vacuum level in the system increases. This is illustrated in the second plot from the top. The relevant switching thresholds are also shown in this plot. As soon as the vacuum level exceeds the threshold value H2, the *Part present signal (H2)* is activated. The same applies to the *In control range signal (H1)* as soon as the switching threshold H1 is reached. Both are shown in Fig. 16 in the third plot from the top. The output signals are often abbreviated to the switching thresholds, since they are directly dependent on them. In contrast to the *part present signal*, the *In control range signal* is only switched to passive as soon as the hysteresis calculated from H1-h1 is undercut. This happens either by suddenly occurring errors or by the blow off signal. Another interesting parameter besides the vacuum level is the current power of the vacuum generator. This is shown in the lowest plot of Fig. 16. It is easy to see when the vacuum generator is active and when the air saving control switches it to passive mode. In the following sections, the use cases for the gripping system will be explained in more detail on the basis of the fundamentals described above.

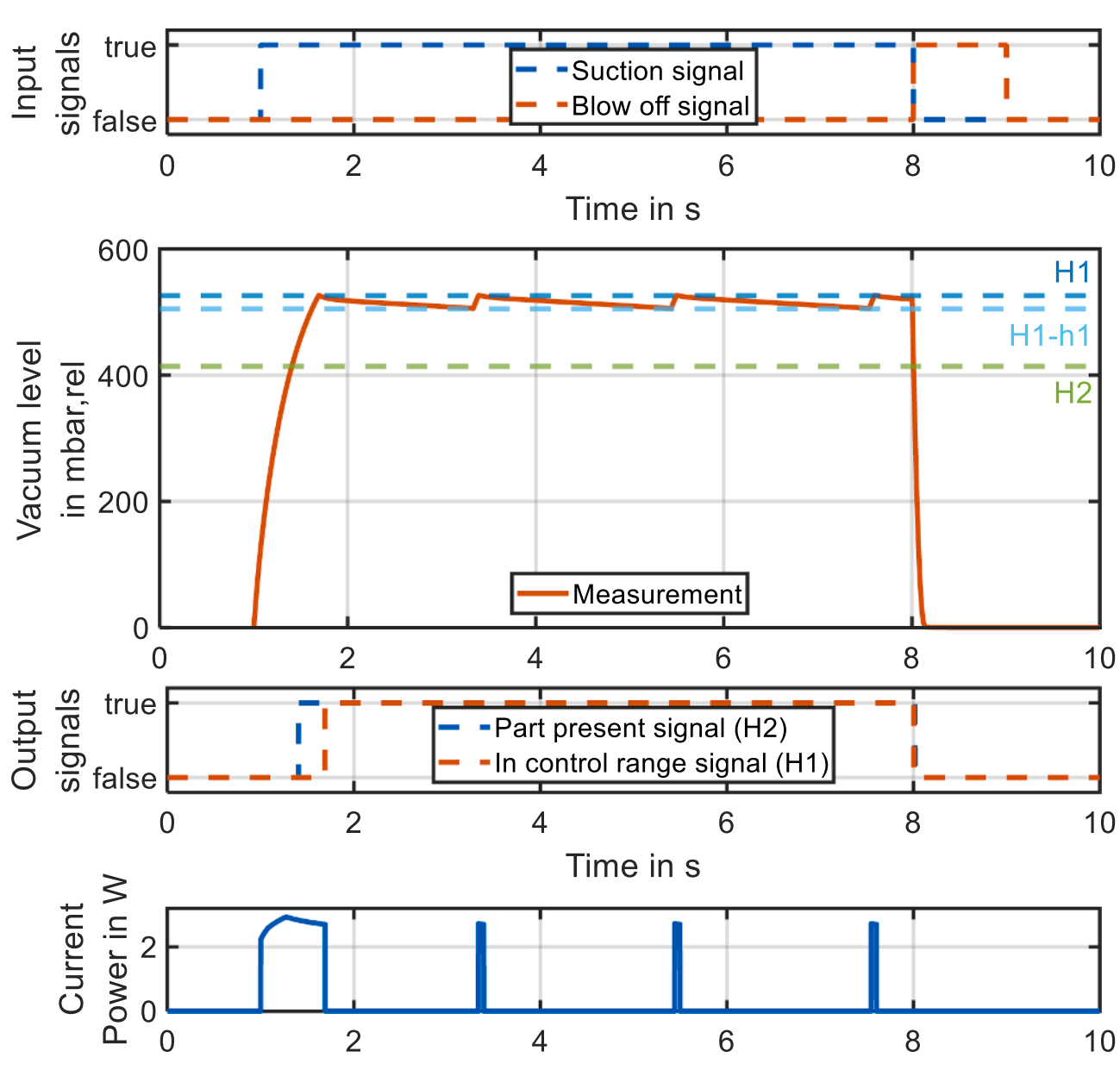

**Fig. 16.** Relevant input, output and internal signals of vacuum gripping system.

### 4.3. Use case 1: product design

A major challenge for component manufacturers in the field of industrial automation is the selection of ideal solutions out of their mostly broad product portfolio. A lot of different combinations can solve the customers' problem with different benefits and drawbacks. Currently, this job is usually done by sales people advising customers mostly based on expert knowledge and experience. The challenge to always select the ideal solution just based on expert knowledge and experience without any testing is nearly impossible with a constantly increasing variety.

In this area the DT can benefit greatly enhancing already existing optimizing algorithms with knowledge about the component's behavior. Through this construct, optimal solutions can be generated automatically with a strongly varying complexity. The challenge addressed also exists in the field of vacuum handling systems. The product portfolio consists of different components with varying properties such as hoses with different lengths and diameters or vacuum generators based on pneumatic or electric energy. The customer is often looking for a solution for his handling task. This handling task is mostly influenced by parameters describing the handling object and the moving process. The number of parameters can vary depending on the level of detail of the analysis and the application. For this use case, the parameters from Table 1 are used.

A central parameter for vacuum gripping systems is the vacuum level. Many system properties such as evacuation time or energy consumption can be derived from it. To get the required vacuum level for the handling task, the needed holding force is calculated first using Eq. 1. [61].

$$F_H = (g + a) * \frac{m_O * S}{\mu} \tag{1}$$

The required vacuum level can be derived from the holding force with the relationship from Eq. 2. The active area of the gripping system results from the suction cup area and the number of suction cups:

$$\Delta p = \frac{F_H}{A} = \frac{F_H}{n * \pi * (d/2)^2} \tag{2}$$

This equation combined with Eq. 1 and the values from Table 1 results in a minimum vacuum of 414 mbar,rel.

$$\Delta p \geq \left(9,81\frac{m}{s^2} + 5\frac{m}{s^2}\right) * \frac{0,15kg * 3}{0,5 * 3 * \pi * \left(\frac{11,7mm}{2}\right)^2} \geq 413,25mbar \tag{3}$$

Vacuum generators, often use integers as switching thresholds. For this reason, 414 mbar is assumed as the minimum vacuum value. The upper switching threshold is often set to 100 mbar above this minimum vacuum value. With these requirements, the next step is to combine components into a system and determine its performance. In this case, the portfolio of J. Schmalz GmbH is used. To reduce complexity, only the selection of the vacuum generator is shown. The gripping system as such was defined manually in advance. However, the concept can be extended to all components of vacuum gripping systems through adding additional complexity and implementation effort.

For the vacuum generators, a manual preselection is made in the use case to increase the comprehensibility. For other use cases, the preselection step can be omitted and an automatic selection from all components can be performed. The preselected components with the parameters relevant for this use case are shown in Table 2.

Similar to the parameters describing the handling process, there are a number of different Key-Performance-Indicators (KPIs) for describing the performance of a vacuum gripping system, depending on the detail of analysis and the application. However, frequently used KPIs are the acquisition cost, the energy consumption for one cycle, and the active cycle time, which is divided into the evacuation time and the blow-off time. With the above parameters for the process, the selected gripping system and the preselected vacuum generators, the automatic creation of the DT is triggered. The models created in this way can then be automatically evaluated with regard to the KPIs mentioned. To explain this process in more detail, the individual steps are described below. The vacuum curve is central to vacuum gripping systems, shown for the different vacuum generators from Table 2 in Fig. 17 on the top.

The lower section of this figure shows the energy consumption for the different vacuum generators over time. From these curves, the KPIs energy demand per cycle and evacuation times can be determined, shown in Fig. 18. It would also be conceivable to use the DTs created in this way for an assistance system that helps sales engineers to design optimal systems. For them, not only the KPIs, but also the values over time matter, as they contain a lot more information they can use to select the ideal solution for the customer. Analyzing the KPIs shown in Fig. 18, it can be easily seen, that the two electric vacuum generators offer major energy advantages over the pneumatic vacuum generators in this application scenario. In terms of cycle times, all the vacuum generators meet the required 800 ms. For this reason, the ECBPMi is selected because it offers major advantages in terms of acquisition costs compared to the ECBPi. Compared to the other three vacuum generators, the ECBPi was not used at the front of the robot, but was placed next to it, since the weight exceeds the maximum load capacity of the robot. A hose with a length of 750 mm and an optimized inner diameter of 3 mm was used for the vacuum supply in the calculation above.

However, the hose diameter as an example shows well the hidden potential in the use of such behavior models from the DT for the design of entire systems. As an example, different hose diameters were investigated with regard to their influence on the vacuum level at the suction cup, at the vacuum generator and the energy consumption per cycle, shown in Fig. 19.

**Table 1**
Use case relevant parameters for handling process [62].

| Parameter | Variable name | Value |
|---|---|---|
| Object weight | $m_O$ | 0,15 kg |
| Object surface | - | Smooth, without holes and no soak through |
| Load case | - | 3 |
| Maximum acceleration | $a$ | 5 m/s^2 |
| Maximum cycle time | - | 800 ms |
| Load capacity of the KUKA youBot | $m_{max}$ | 0,5 kg |
| Safety factor | $S$ | 3 |
| Friction coefficient | $\mu$ | 0,5 |
| Suction cup diameter | $g$ | 11,7 mm |
| Number of suction cups | $n$ | 3 |
| Gravity | $g$ | 9,81 m/s^2 |

**Table 2**
Relevant parameters of preselected vacuum generators.

| | ECBPMi | ECBPi | SCPMc 03 | SCPMc 05 |
|---|---|---|---|---|
| Max. suction capacity | 1,6 l/min | 12 l/min | 2,2 l/min | 7,5 l/min |
| Max. vacuum | 600 mbar | 750 mbar | 870 mbar | 870 mbar |
| Max. input power | 3 W | 13 W | 3,5 l/min | 9 l/min |
| Drop off principle | Valve | Valve | Blow Off | Blow Off |
| Costs | 995 € | 2.260€ | 290€ | 295€ |
| Weight | 240 g | 775 g | 65 g | 70 g |
| Positioning | On gripper | Besides robot | On gripper | On gripper |

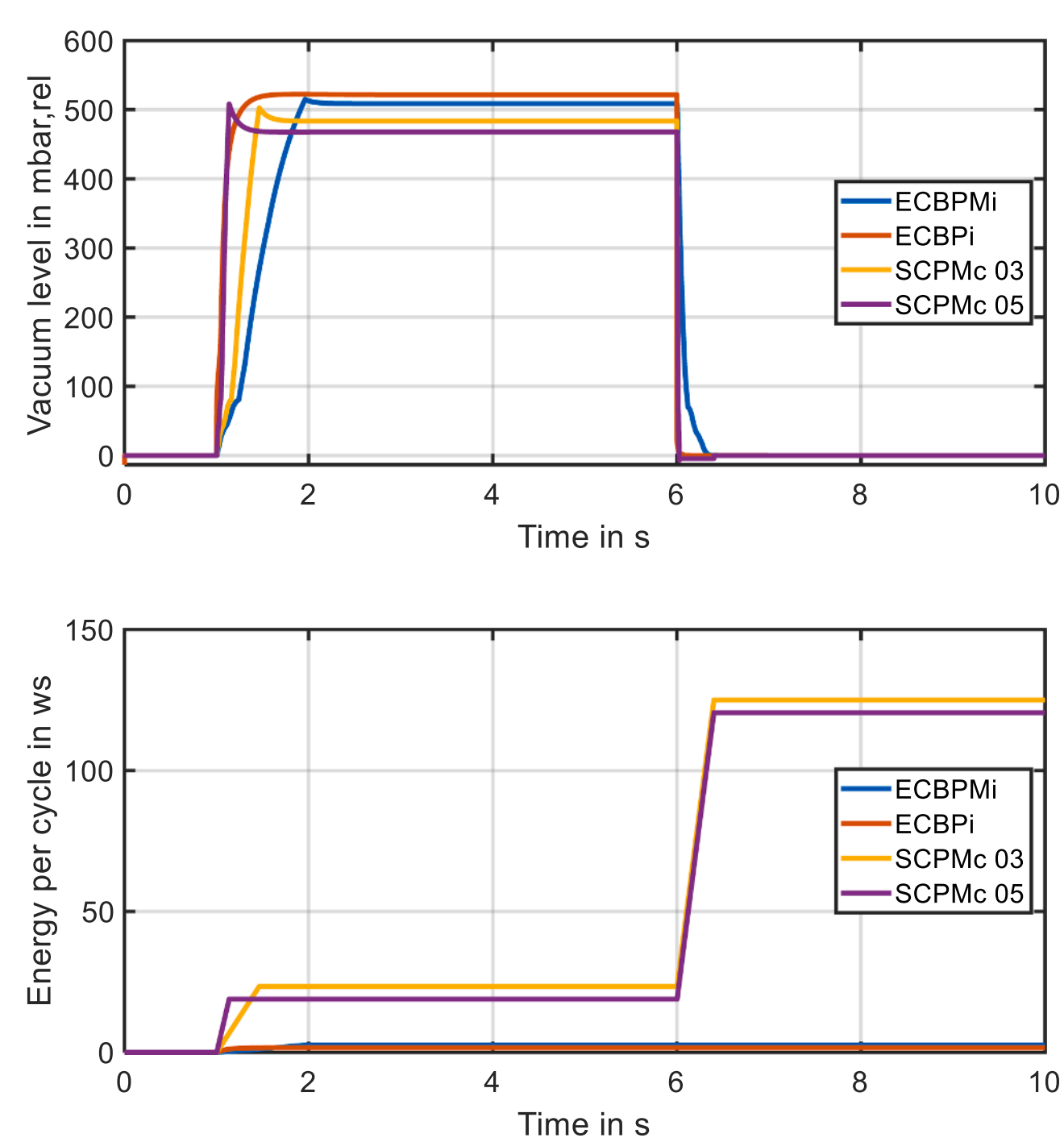

**Fig. 17.** Vacuum level and energy over time for different vacuum generators of use case.

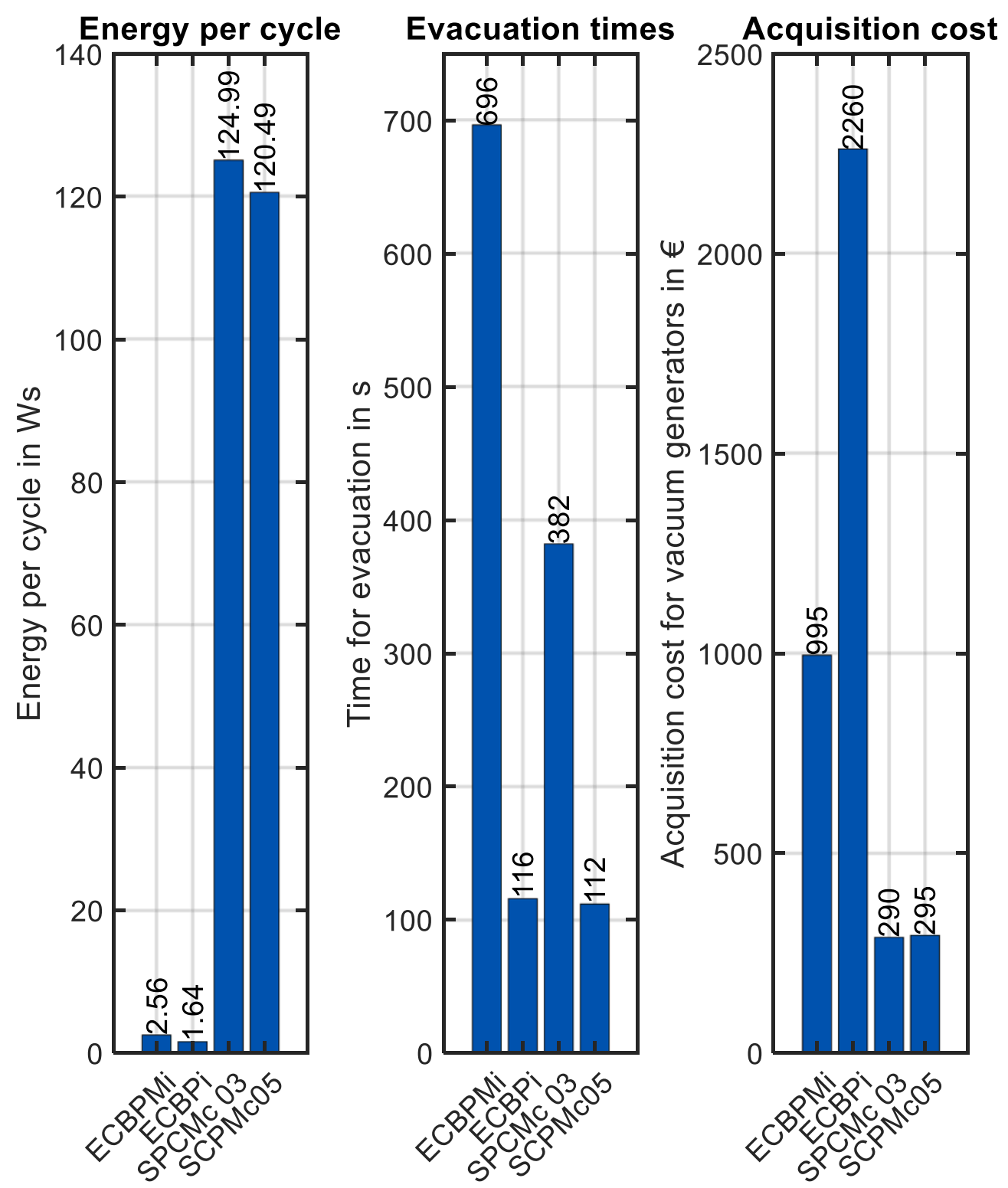

**Fig. 18.** Energy consumption, cycle times and acquisition costs for different vacuum generators.

It can be seen easily that there is an optimum in terms of cycle time and energy consumption at 3 mm. Such an optimum can be found not only manually but also with appropriate algorithms. The system design of this section results in the system for the further use cases. This consists of three suction cups, a 3D-printed manifold structure with internal vacuum guide and the ECBPMi as vacuum generator. This gripping system is shown in Fig. 20.

### 4.4. Use case 2: design validation

After the design of a system as shown in Section 4.3, currently physical systems are often built for validation. This can be done virtually too by using behavior models in the DT avoiding the costly and time-consuming construction of real systems. During such validations, different influencing parameters (Fig. 15) and their effect on the functionality of the system are checked. For vacuum gripping systems, an influencing parameter with great significance is the leakage of the system. In a faultless system, leakage often originates from the handling object or through the suction lip. Since smooth metal parts are considered in the application scenario under consideration, leakage through the component can be ruled out. The same applies to the suction lip on new suction cups, which seals well against a smooth surface like metal parts. However, the suction cup changes during lifetime due to wear. As a result, the suction lip is worn and leakage at the point between the suction cup and the handling object increases. The influence of this leakage is analyzed in this use case.

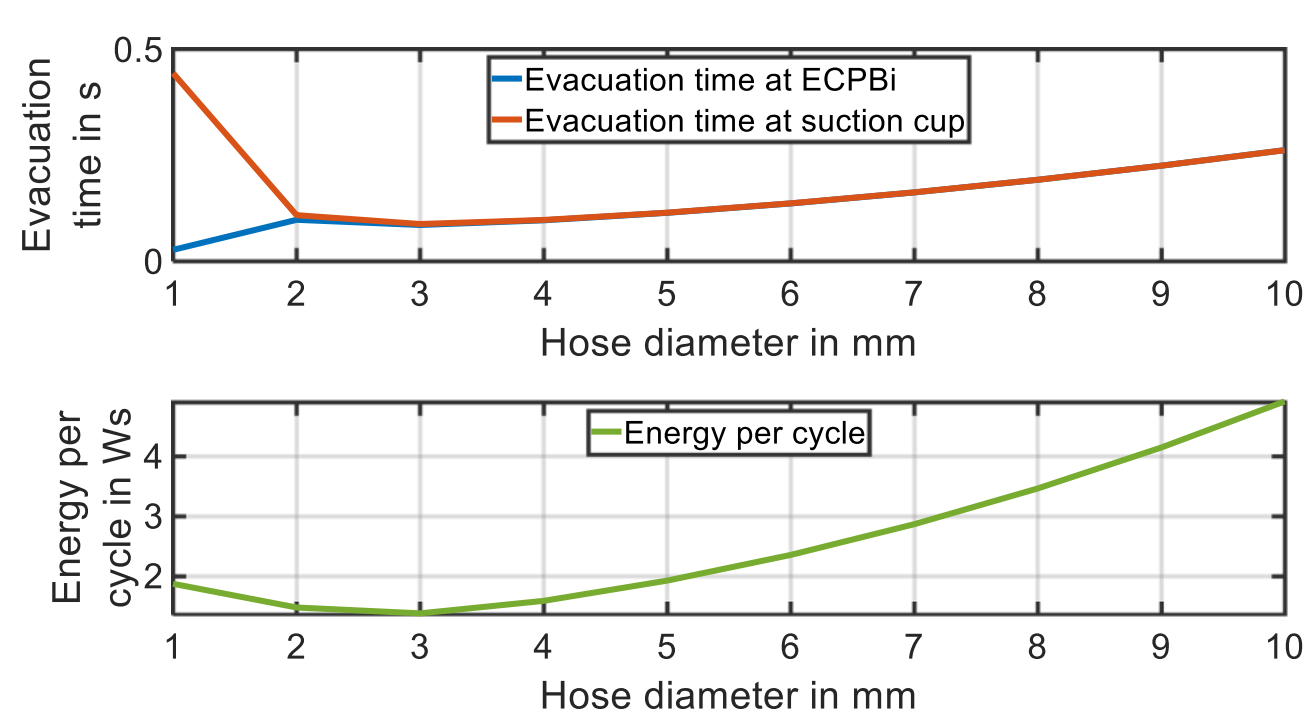

**Fig. 19.** Evacuation time and energy per cycle over hose diameter for the ECBPi.

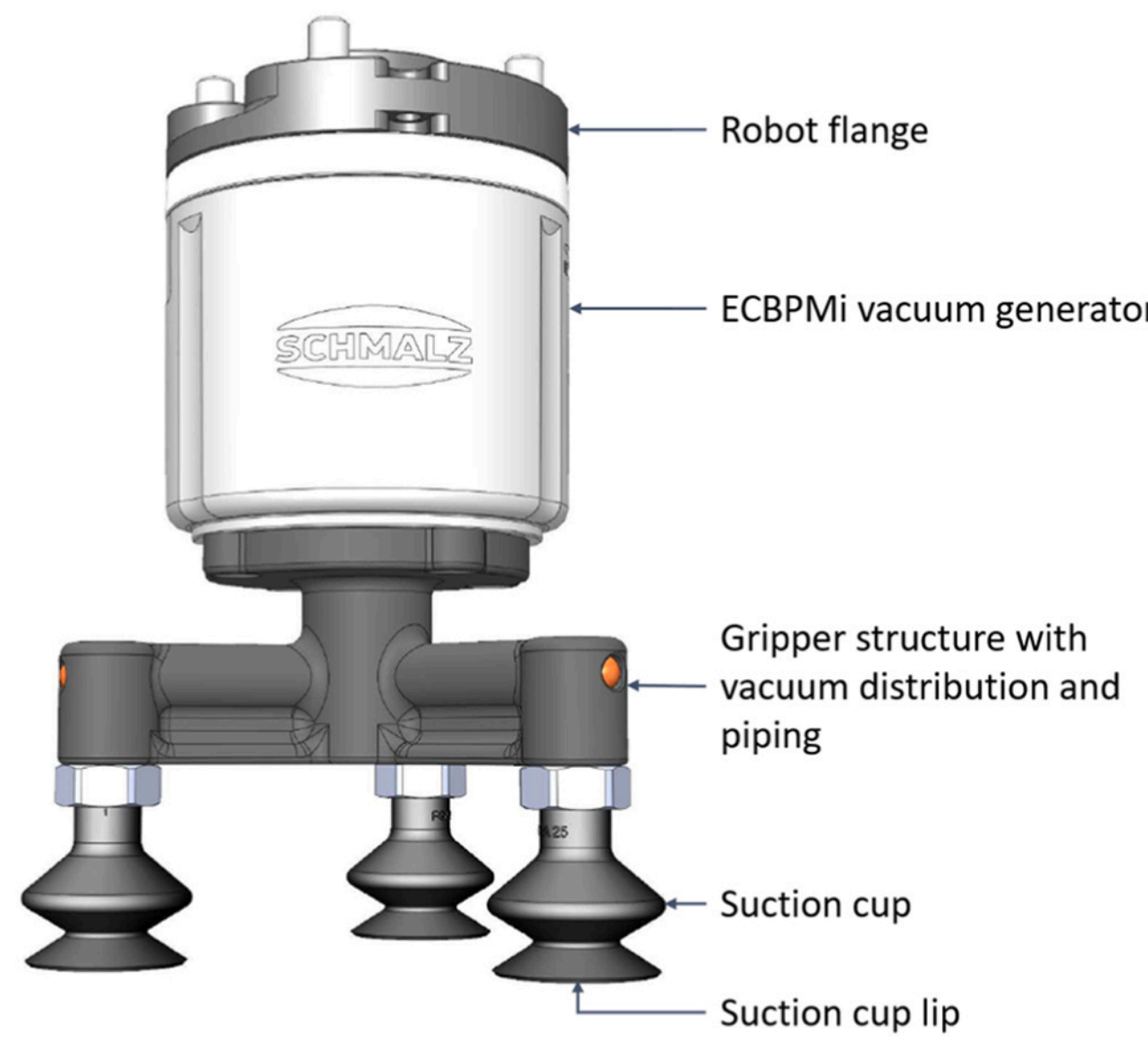

**Fig. 20.** Designed vacuum gripping system for the other use cases.

Another influencing parameter in a modular cyber-physical production system is the weight of the handling object. It can vary, for example, when the handling object is changed or adapted. The weight has a direct connection to the influencing parameter leakage, since the leakage is dependent on the vacuum level. Since a higher handling object weight requires a higher vacuum level, leakage and handling object weight influence each other directly.

The influence of different handling object weights on the evacuation time as well as the energy consumption at different leakage diameters is shown in Fig. 21.

This quickly shows that the vacuum gripping system with the designed 0.15 kg can tolerate a certain amount of leakage. The higher the handling object weight, the less leakage can be tolerated. With regard to energy consumption, it can be stated that both, additional weight and additional leakage result in increased energy consumption. The limit function of the correlation between leakage and handling object weight can be easily determined too. Therefore, the authors use the automatically created and parameterized behavior models from the DT with existing analysis algorithms. It indicates the parameter combination where the vacuum generator can still handle a specific combination of weight and leakage. This area is shown in green in Fig. 22.

So, depending on the current leakage in the vacuum gripping system, different maximum weights for the handling object can be handled. However, this is only the case if the other parameters such as safety factor, robot acceleration and friction coefficient remain constant. Such information can be either presented to the customer of the system directly or provided via the DT. This would enable the system overlaying the vacuum gripping system to use this dependency to plan production or adapt parameters.

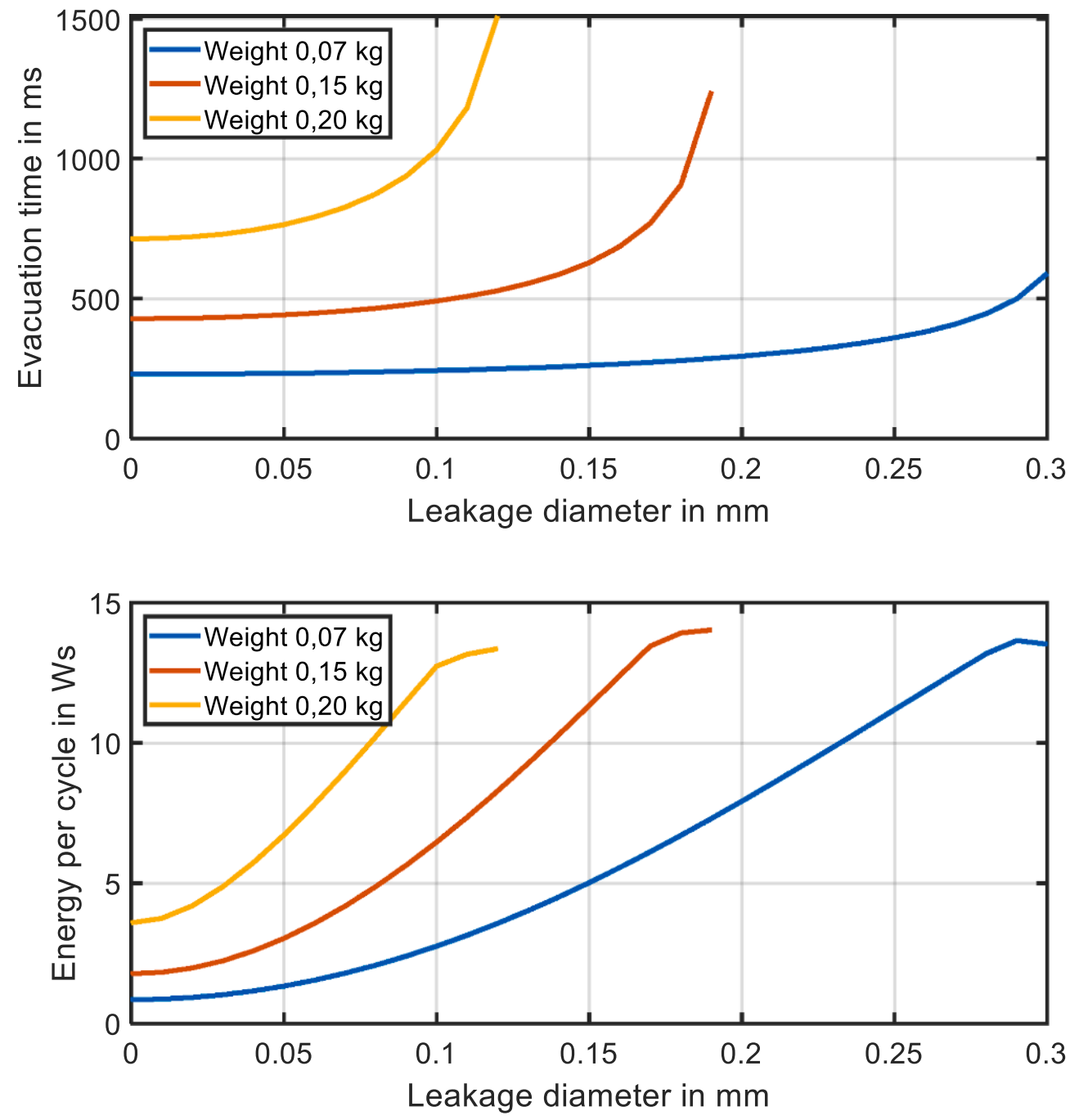

**Fig. 21.** Evacuation time and energy over leakage diameter for ECBPMi.

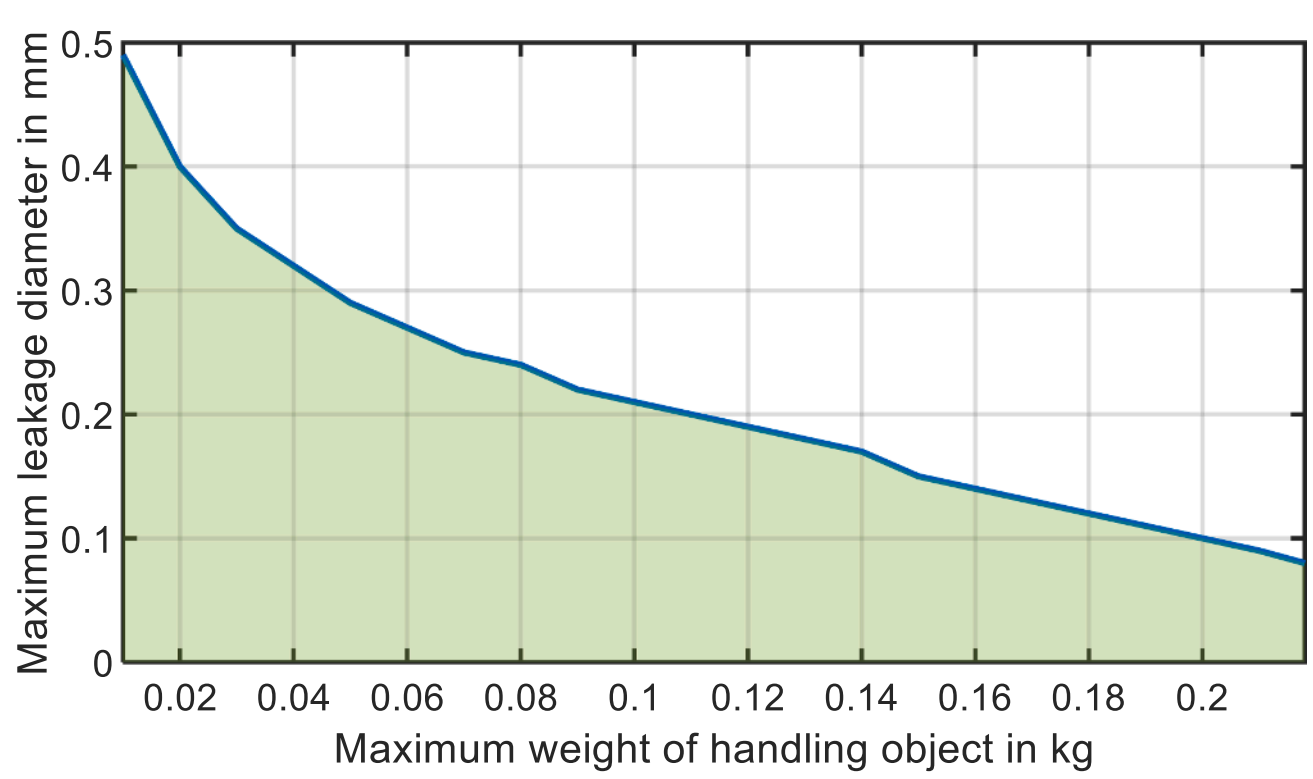

**Fig. 22.** Limit for leakage diameter over handling object weight for ECBPMi.

### 4.5. Use case 3: instance specific data

Ahead of production, the components under consideration are still virtual components. The parameters in the corresponding models used in the DT are object-specific values. These originate either from previous measurement series or other more in-depth simulations from the *physical spatial behavior* level. However, since the DT can be assigned individually for each instance of a component, it allows instance-specific data to be assigned. This can be done at the end of the production of a component for example. During the end-of-line test, the recorded values are stored in the DT of the corresponding component. An exemplary usage of these instance-specific data is to reduce the simulation-to-reality gap. The data from a measurement of the ECBPMi with a solid tank is shown in Fig. 23 similar to [23].

The measured vacuum curve together with the simulated vacuum curve with instance-specific and object-specific data are shown in the upper part. The measurement was repeated ten times to reduce the influence of statistical errors. In addition to the blue curve, which represents the mean value of the ten measurements, an envelope curve is shown too, representing the minimum and maximum values of the ten iterations. Since the scatter of the measurements is extremely small, this envelope curve is not visible in the upper representation. In order to give a more detailed explanation, a section of the upper representation is shown with adapted axes ranges in the middle of Fig. 23. This shows a slight scatter between the minimum and maximum values of a maximum of 8.55 mbar. The bottom figure illustrates the deviation between instance and object-specific data. It shows the absolute deviation in mbar between simulated and measured behavior for the two different data sources. It clearly shows the reduction of the deviation between measured and simulated behavior by using instance-specific data for simulations in the DT.

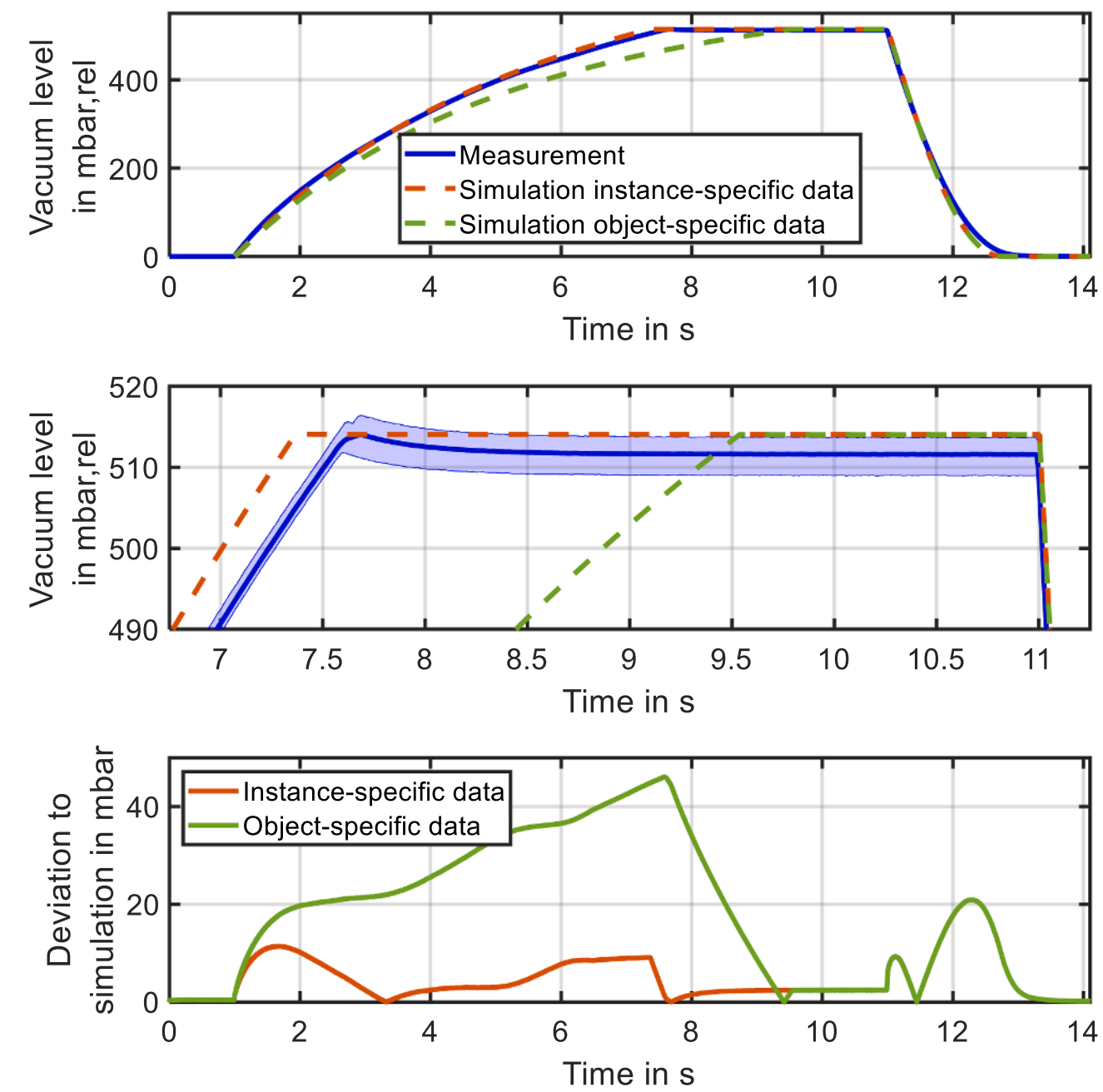

**Fig. 23.** Improvement of simulation accuracy through instance-specific data.

Besides the described case of increased simulation accuracy, instance-specific data can also be used for other use cases. An example would be the provision of such data as a service for adjusting the accuracy of vacuum sensors [63]. In addition to these two examples, further value can be added using instance-specific data.

### 4.6. Use case 4: virtual commissioning

In the virtual commissioning use case, the transfer of the real component from the component manufacturer to the machine operator takes place. In parallel to the physical exchange of components, DTs are also to be transferred from the component manufacturer to the machine manufacturer for efficient virtual commissioning. Here, different components are brought together increasing the model scope (from component to station). In addition to the modeling scope, other model properties such as the modeling depth or the modeling width often change too. If, for example, the component manufacturer paid great attention to fluidics, this no longer plays an important role for the machine manufacturer. In this specific case, the virtual commissioning primarily considers the physical behavior of the robot arm in combination with the vacuum gripping system interacting with the handling objects. The execution of software-in-the-loop and hardware-in-the-loop cases can take place. For example, the implemented control code and thus the trajectories of the robot are checked in the simulation using GAZEBO (see Fig. 24).

In addition, error cases, such as unsuccessful gripping of the handling object can be easily tested without complex test setups and the risk of destroying either the robot-arm-gripping-system or the handling object.

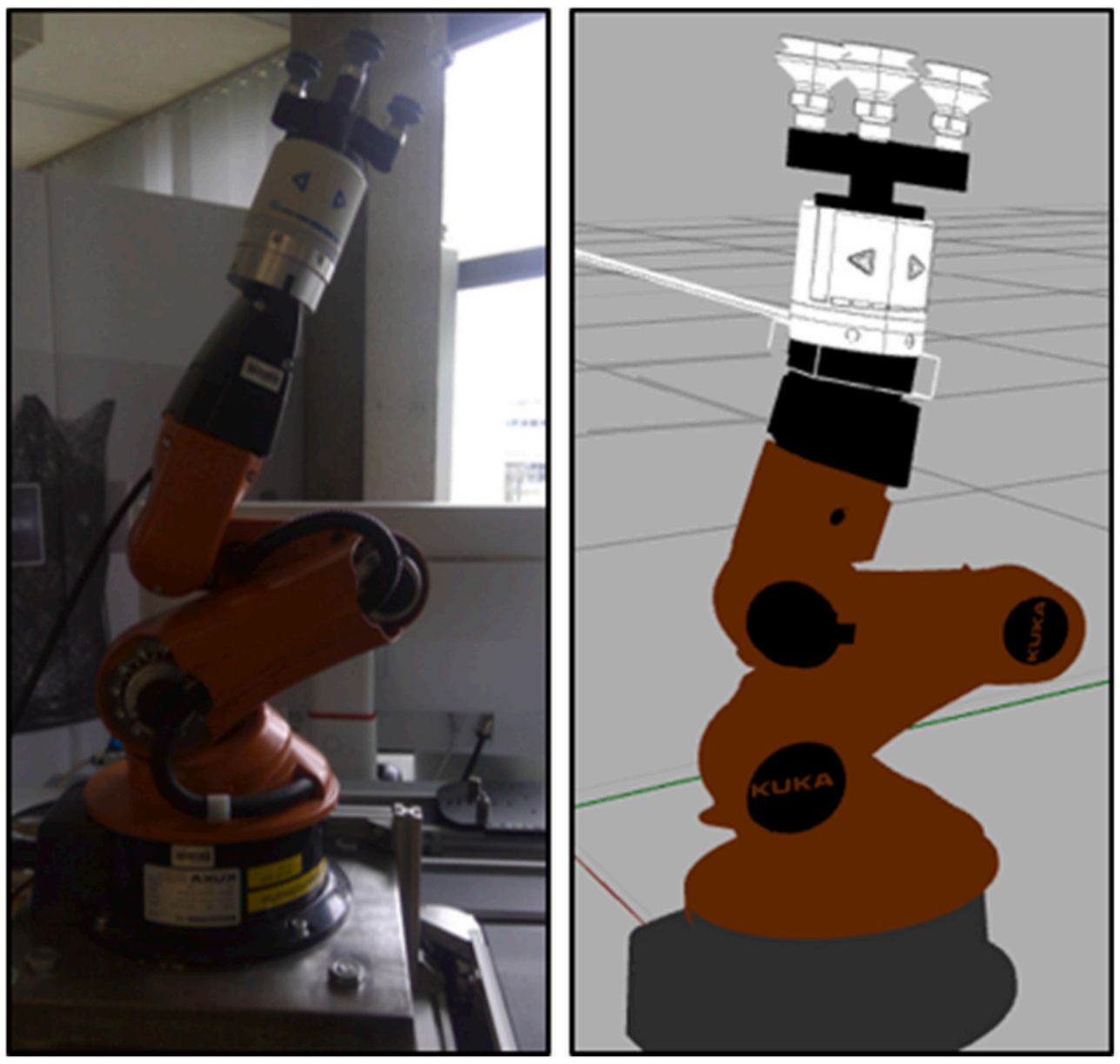

**Fig. 24.** Real Robot-arm-Gripping-System (left) and Simulation (right) in GAZEBO.

### *4.7. Use case 5: operation-parallel simulation*

In the operational phase, machine operators are often faced with the challenge of understanding complex production systems they did not develop on their own. Therefore, they mostly do not have distinctive expert knowledge about the systems. Their goal is to operate their machines as efficient as possible. Ever shorter product life cycle phases require effective changeover and fast start-up curves towards optimal machine operation. Thus, continuous virtual commissioning may become necessary more frequently during the operational phase for new products on an existing machine. In this case, the operation-parallel simulation can help to explain the complex system to the employees through training. Furthermore, remote monitoring is made possible, for example, or non-measurable process parameters can be simulated. From the authors' point of view, the operation-parallel simulation with the help of behavior models can be the basis for further applications such as diagnosis, prognosis or optimization.

In this context, the operation-parallel simulation can initially serve as an abstracted material flow monitoring, in which the individual coupled component models of the cyber-physical production system (e.g. AGV, conveyor belt, robot and gripping system) map their cycle time in order to enable cycle-based monitoring. This can be represented by *discrete-time behavior* models according to [26]. This enables, for example, the comparison of the actual and target planning and thus the monitoring of the overall cycle time and indirectly the Overall Equipment Effectiveness (OEE). The simulated cycle time of the overall system must be equal to the real cycle time. In the event of deviations from the real cycle time, these can then be assigned to individual components and a diagnostic application can search for a cause more efficiently.

The prognosis of the system behavior is another application of the operation-parallel simulation and can provide values for predictive maintenance, for example. Here, the simulation is required to simulate the systems behavior faster or equal fast compared to the real behavior. The values of the prognosis application can then in turn be used for the optimization application, e.g. for improved process parameters. The machine operator, as an example, can increase the travel speed of the robot to reduce the total cycle time. Increasing the travel speed increases the forces to be applied by the vacuum gripping system to keep the handling object on the robot's path. If the maximum force of the vacuum gripping system in the current parameterization is not sufficient, the machine operator can increase the suction force by increasing the vacuum level before implementing this scenario to the real system.

As mentioned above, the execution times of behavior models for operation-parallel simulations play an important role depending on the application (diagnosis, prognosis, optimization). Fig. 25 shows the execution times for the simulation and the compilation and for just the simulation of the different modelling depths of the gripping system with the ECBPMi.

For this purpose, the models were executed in MATLAB by a script for regular and fast restart mode. Using the fast restart mode, the compilation of the model is omitted and the pure simulation time can be determined. The models were executed ten times and an average value was calculated from the determined times. The simulated time period was 9 s. The system used is a commercially available computer with an Intel Xeon W2123 processor and 32 gigabytes of RAM. Prior to the executions of the models, all other programs on the computer were actively closed.

These times are decisive for the selection of the modeling depth depending on the application and its requirements. The percentage comparison in grey clearly shows the differences between the individual modelling depths and gives a perspective on the difference between modeling depths for much larger systems. The calculation times between *discrete behavior*, *discrete temporal behavior* and *continuous simplified physical behavior* increase slightly. The jump between *continuous simplified physical behavior* and *physical non-spatial behavior*, on the other hand, is more striking. From the author's point of view, the computing time for *discrete behavior* to *continuous simplified physical behavior* only becomes relevant if very large simulations are involved, for example of all components of the cyber-physical production system in an overall simulation, or if the real-time requirements for the simulations are very high. Providing such meta-information about the behavior models is an important part of model comprehension in an iDT.

Fig. 26 is used to illustrate the differences over the course of time between the modeling depths using the vacuum level of the vacuum gripping system. In this plot, the vacuum curve over the time of the described cycle is shown for the ECBPMi at the different modeling depths. In order to highlight the respective deviation from reality, the measured value is also shown in blue. For a more detailed explanation of the curves, please refer to a similar representation in [26].

For the forecast case described above, the fastest execution through *discrete behavior* could be decisive in order to enable a pre-simulation of discrete states. Otherwise, the difference in computing time between the models is almost negligible. For the jump between *continuous simplified physical behavior* and *physical non-spatial behavior*, however, it must be decided whether the use case requires such a high accuracy of the simulation that the significantly increased computing time is justified.

In this context, *discrete behavior* models can also provide relevant information about any failure cases or increasing wear. An example for this is the switching threshold H1 at the example gripping system. This is used to stop the evacuation for the vacuum gripping system at a certain value and to switch to vacuum control. For the ECBPMi, this switching threshold is 20 mbar above the switching threshold H2 relevant for the process, which serves as a start signal for the robot. Thus, if the switching threshold H1 is no longer reached, the process continues anyway. Such a scenario is shown in Fig. 27. In the upper section, the vacuum curve is shown over a period of 1000 s. As the time increases, the leakage in the system increases artificially fast. The time is deliberately chosen to be so short for reasons of simplicity.

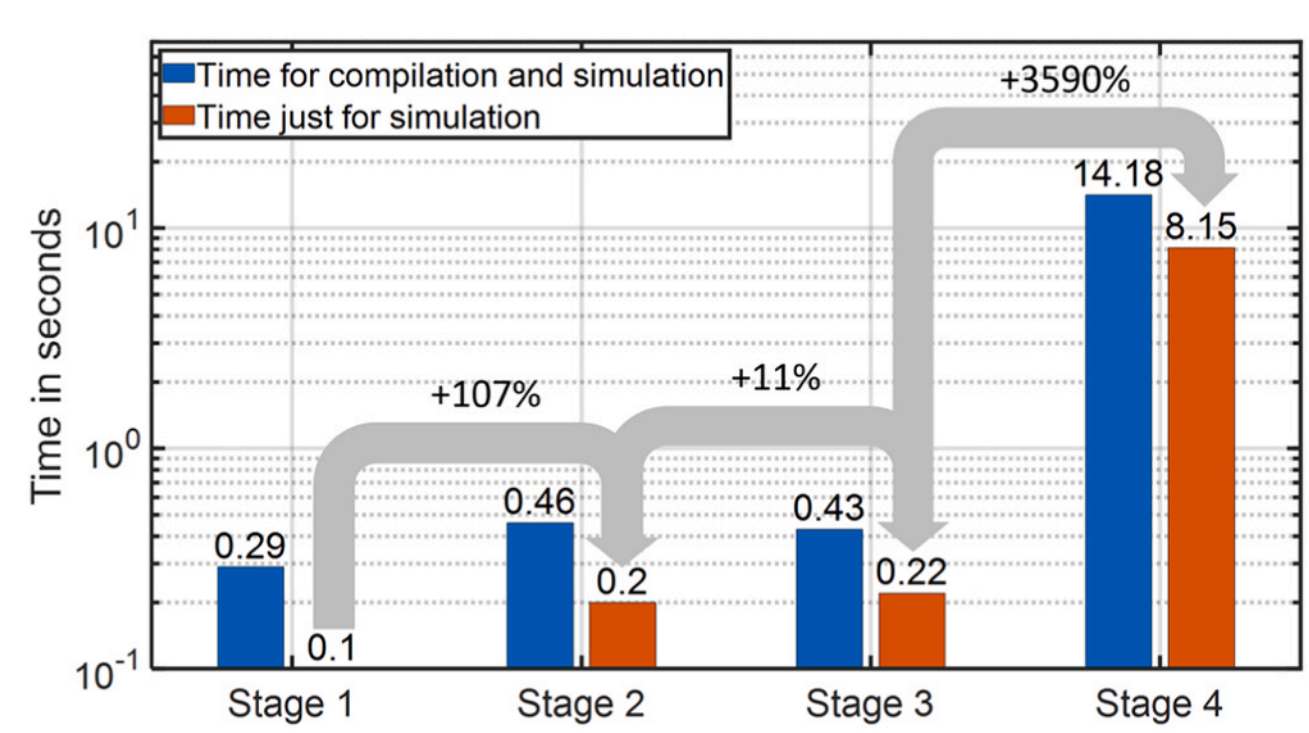

**Fig. 25.** Time for compilation and simulation of behavior models of different modeling depths.

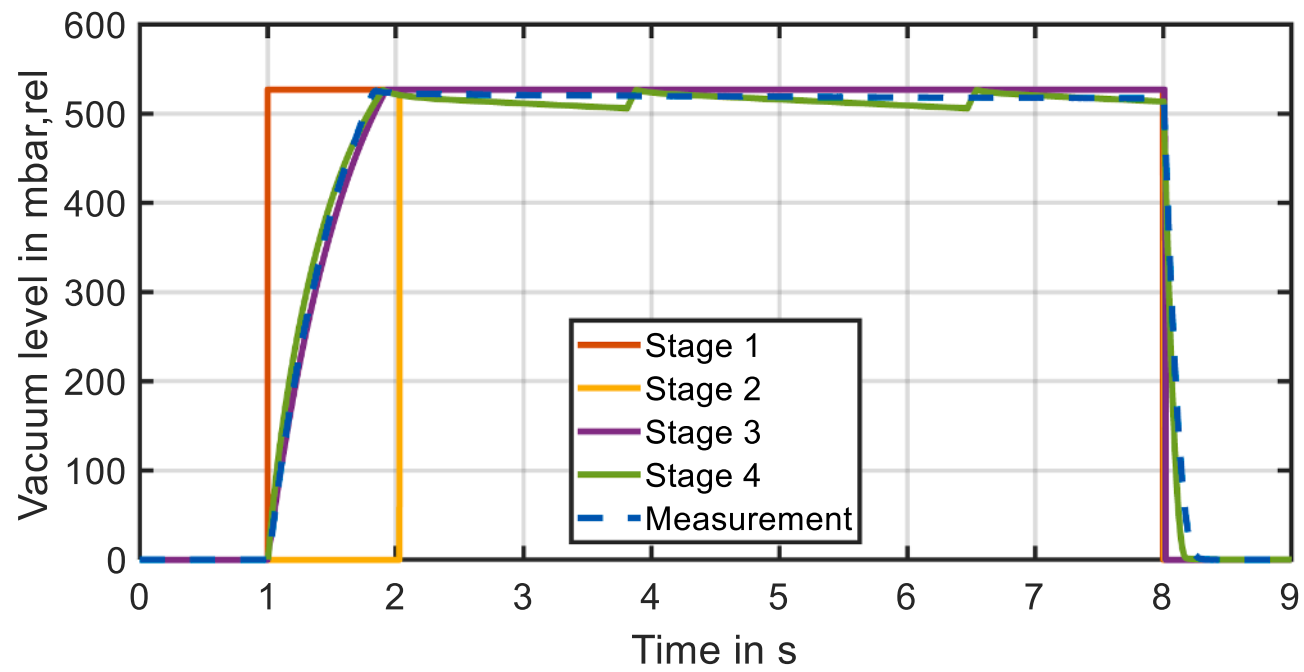

**Fig. 26.** Comparison of measurement and simulation with different modelling depths without leakage.

In real scenarios, the time span would be a lot longer. In Fig. 27 it is easy to see that the maximum vacuum in the system decreases with increasing leakage. The H1 signal is also shown in the middle of the figure. The signal of the discrete behavior model is shown in blue and the signal of the real process is shown in orange. It is easy to see that the switching threshold H1 is no longer reached in reality after about 600 s. However, the discrete behavior model still outputs a signal H1. From this gap, an increasing leakage can be recognized as a symptom of advanced wear, for example of the suction lip. However, the process itself can still continue. Only when the switching threshold H2 can no longer be reached, the robot is no longer able to starts its movement process leading to an overall standstill of the entire process. This is shown in the lower part of the figure. It can be seen that this state occurs after about 800 s

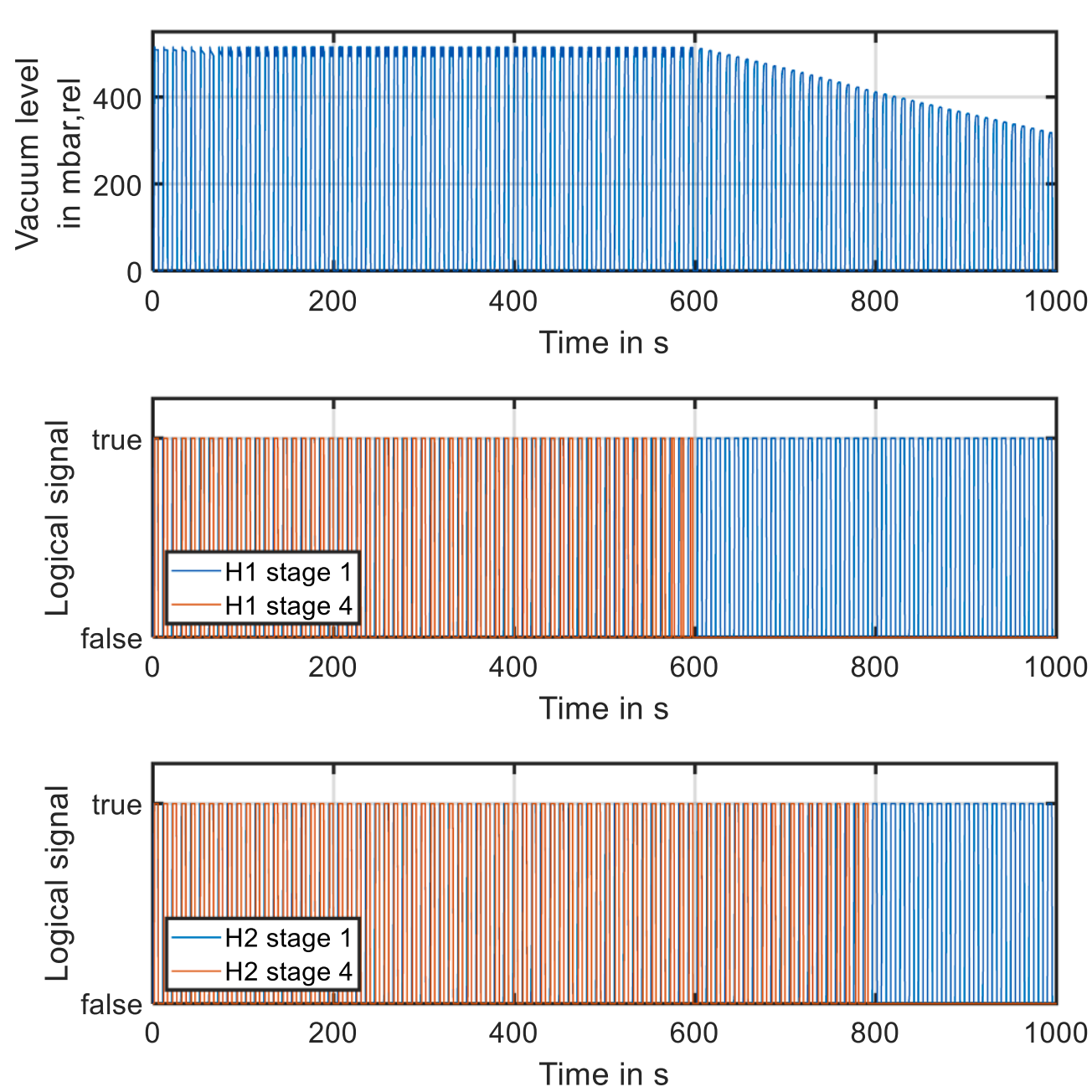

**Fig. 27.** Influence of artificially increasing leakage at vacuum level, switching threshold H1 and H2.

### 4.8. *Use case 6: automated failure detection*

As described in the previous use case (Section 4.7), a diagnostic process can be started in the event of a deviation in the operation-parallel simulation in order to identify the reasons for the deviation. For this purpose, the behavior models with a higher modeling depth can be used, as they consider an extended system boundary by adding influencing parameters mentioned in Fig. 15 (e.g. hose diameter, leakage, ambient pressure). Thus, changes in the environment or in the asset that cause a deviation can be considered by certain parameters in higher modelling depths. The behavior models can be used as a 'digital playground' to identify these parameters in terms of diagnostics without having to perform tests on the real asset. In order to trigger a diagnostic use case, a variety of possible changes can be introduced at the cyber-physical production system. For this paper, the leakage scenario at the suction cups was introduced as a change at the real gripping system to trigger a deviation and thus the trigger for the model adaption.

If these different models are available to the machine operator, he still faces the challenge of using them. As described in Section 2.3, software and model knowledge are required to identify suitable models. Furthermore, knowledge is necessary for parameter adaption in order to fit them into the model with realistic lower and upper limits. Assuming that the machine operator has a deviation for a component of the modular production system detected using the operation-parallel simulation and has the required software and model knowledge too, a manual model adaption can be carried out for the diagnostic application.

Two cases can be distinguished. In the best case, a diagnostic expert with sufficient knowledge of the simulations used is directly available to start the troubleshooting. Relevant parameters for localizing the source of the error are considered right at the beginning of the diagnosis and the result can be verified after a short time. In the worst-case scenario, no expert knowledge of the simulation models is available. First, the exact model to be used for the diagnosis must be identified before the relevant parameters can be searched for and validated. However, it is optimistic to assume that machine operators have expertise for all models of a modular production system they use. In order to find the suitable model for closing the reality gap, the procedure is as follows.

After detection of the deviation, the *Plan* step is triggered. Since this experiment involves a constant application (operation-parallel simulation) and no new requirement for the application was specified by the user, the actual analysis of the *Plan* step shows that a change has occurred in the gripping system or its environment [24]. The increase of the realism is given as adaption goal. Such an increase is possible by increasing the modelling depth. The structuring of the models in the model pool according to [26] allows an iterative approach for finding the suitable model in the *Do* step. Starting from discrete temporal behavior, continuous simplified physical behavior and physical non-spatial behavior are selected and a parameter adjustment is performed in each case. Fig. 28 shows the results of the *Check* step.

The upper plot shows the vacuum level of the different modeling depths and the real data (blue). Below this, the absolute and the summed deviation over time is shown in order to highlight the significant deviation differences. Only the physical non-spatial behavior (green) can adapt to the real course, since this modeling depth considers the leakage as a parameter. The pointed outliers (middle graph) show the deviation in plot 1 (red circles) and are due to the dynamic process. In this case, a suitable model can be found and switched active in the *Act* step. The adaption approach can therefore first identify the suitable modeling depth with the appropriate parameter set to close the reality gap. In this way, the machine operator can identify problems for the modular production system. In this case, he needs to look for leakage on the gripping system and initiate maintenance. If the machine operator does not have this knowledge, it would also be possible to automatically send the information back to the component manufacturer via an AAS interface. Based on his experience, the component manufacturer can quickly

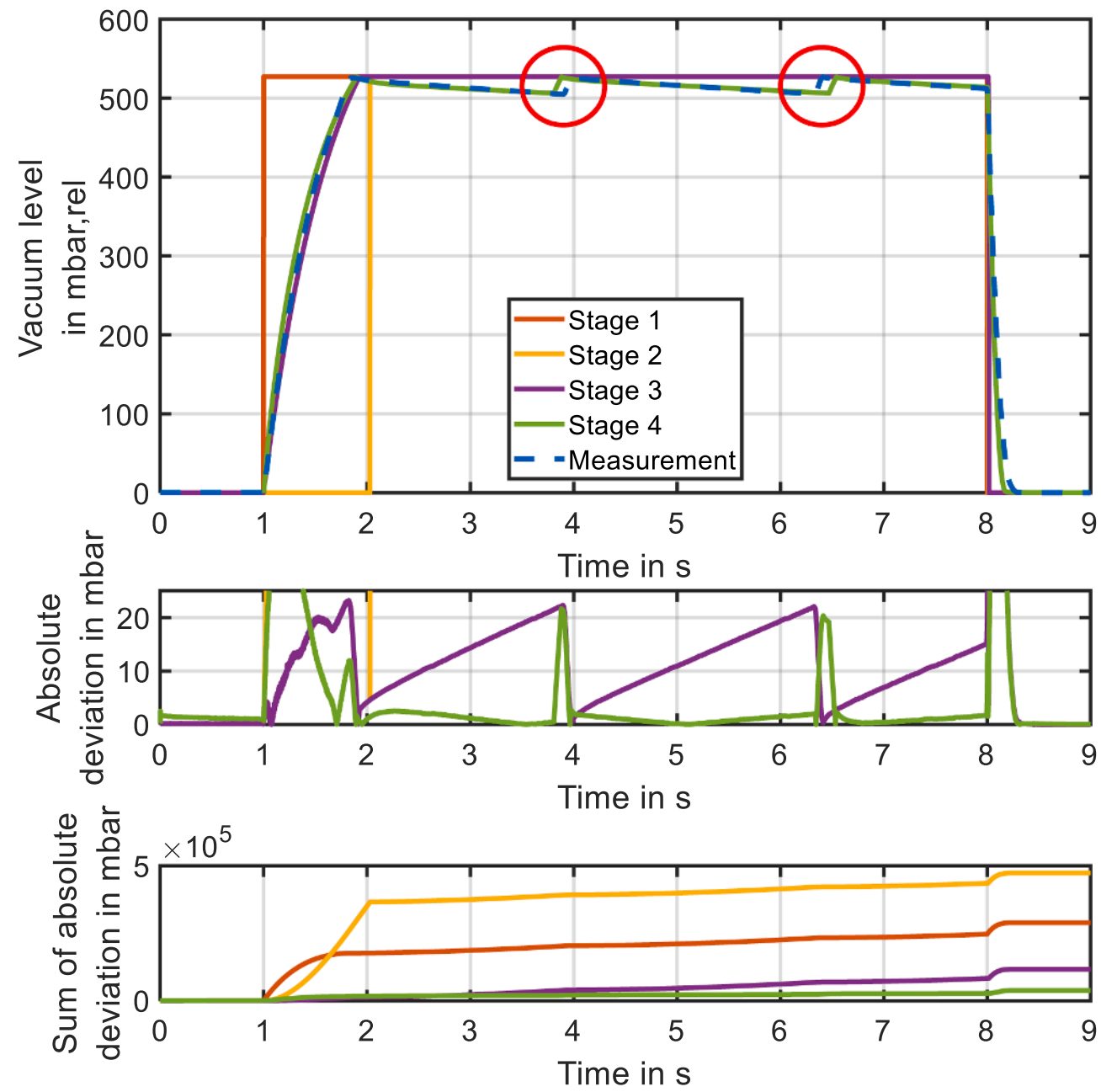

**Fig. 28.** Comparison of measurement and simulation with different modelling depths with leakage.

recognize that the suction cup is probably worn out and can offer a new one as a service.

The manual and automatic procedure of model adaption is the same. As shown in Fig. 29 the model adaption approach presented in Section 3.2 leads to a significantly faster adaption time, since the automatic model selection, parameter adaption and the necessary expert knowledge could be realized by the MAS.

In this scenario, the partial model agents of the individual modeling depths call the parameter estimation function for MATLAB-Simulink via the JAVA-MATLAB interface. Due to the container solution of the agent framework in Jade, the activities in the *Do* step and *Check* step can be executed in parallel. The best-case times of the manual and automatic model adaption show the time needed to run through the modeling depths discrete temporal behavior to physical non-spatial behavior as well as the assumption that the correct parameter set in model depth physical non-spatial behavior is selected as initial values in the first iteration. In the worst case, these ideal initial values do not correspond to the real behavior and the suitable parameter sets must first be found. The error diagnosis scenario clearly shows how time-consuming the manual model adaption is. This is shown here for one component of the overall modular cyber-physical production system. As mentioned earlier, the model pool in the DT will grow rapidly in the future, making manual model fitting economically unattractive. Therefore, automatic

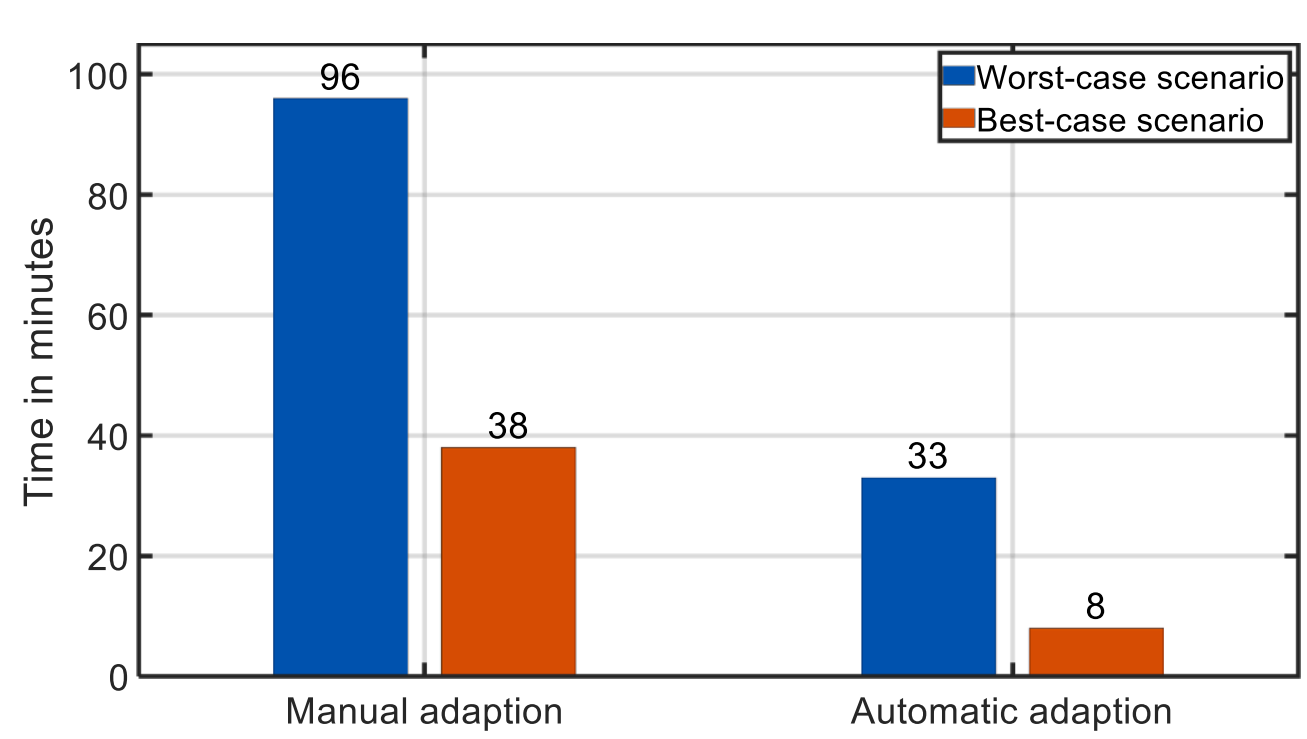

**Fig. 29.** Manual and automated model adaption.

model adaption is needed, enabling the reuse of models in the operational phase.

### 4.9. Use case 7: recycling

In the recycling use case, two fundamentally different applications can be identified. On the one hand, the reuse of a component in its current form in another industrial automation system and, on the other hand, the reuse of the raw materials of a component for the production of new components and systems.

For the first application, the iDT can fundamentally help in understanding how the component works. If the new system in which the component under consideration is to be reused is planned digitally, the different models of the component, which is always kept synchronized, can be integrated in the various machine life cycle phases. This ensures simple and efficient testing of the component's compatibility and actual performance. By using the models from the iDT, the reuse of components is thus possible much more efficiently.

The second application concerns the reuse of the raw materials of a component. An example would be the reuse of the raw materials of a worn-out component such as a vacuum suction cup. This is worn out after a certain number of cycles and must be replaced. A central challenge in the reuse of its raw materials is the exact composition of these. If the exact type of plastic or aluminum used is not known, this often has a negative impact on the reuse rate. This is where the iDT can help by providing information about the asset. Each asset is uniquely identifiable and can be assigned to an iDT. This provides the materials used and their geometric characteristics. In this way, different materials can be easily separated and efficiently reused.

### 4.10. Discussion

For a summarized understanding of the presented use cases, they can be evaluated against a total of seven different criteria. These criteria established by the authors are intended to clarify the experiences and assessments that could be gained with the use cases. The evaluation is not quantitative, but is intended to be a qualitative assessment. The criteria used are explained below. The first two criteria relate to the effort involved in the respective use cases. While the *initial implementation effort* is essentially a one-time effort, the *level of automation* addresses the reduction of recurring manual tasks triggered by the respective use case. Criterion 3 relates to the acceptance of a corresponding use case in the industry. The corresponding benefit such a use case can finally deliver in industry is evaluated by criterion four. Criteria 5 and 6 in turn describe in more detail where the respective *potential benefits* comes from. Is it generated by the *substitution of manual tasks* (criterion 5) or by the creation of *new functionality or services* (criterion 6). The final criterion (criterion 7) in turn describes the status level of the prototypical implementation of the respective use cases by the authors. The individual use cases are evaluated using a blue bar for the respective criterion. The higher the bar, the greater the value of a use case for the criterion, shown in Fig. 30.

The concepts presented in Section 3 mainly serve use cases 1 to 3 as well as 5 and 6. For these, it can be seen that although the concepts involve an increased initial effort, the recurring effort can be significantly reduced, which is reflected in a high level of automation. The industrial relevance of these use cases is still low, but their potential benefit for later use in industry is very high. On the one hand, these benefits stem from the automation of previously manual tasks, such as the automated model creation for use cases 1 and 2 and the automatic model adaption in use case 5. On the other hand, the concepts can offer new services, such as Accuracy-As-A-Service as an example for use case 3 or the automated failure detection in use case 6. Criterion 7 shows that the core use cases of the two concepts have already been implemented to a very high degree in prototypical form. The situation is different for use cases 4 and 7. For virtual commissioning (UC4), this is mainly due to the widespread utilization of this use case in industry, which limits its relevance for research in the view of the authors. The opposite is true for recycling (UC7), which is currently not very widespread but will be central to resilient production in the future. From the authors' point of view, recycling offers huge potential benefits, both in terms of substituting previously manual tasks and in terms of creating new functionality and services. If the DT created and adapted by the two concepts presented is used, there is an initial and recurring effort, but this can be significantly reduced by using the models from the DT. This use case will be further conceptualized and elaborated in future work.

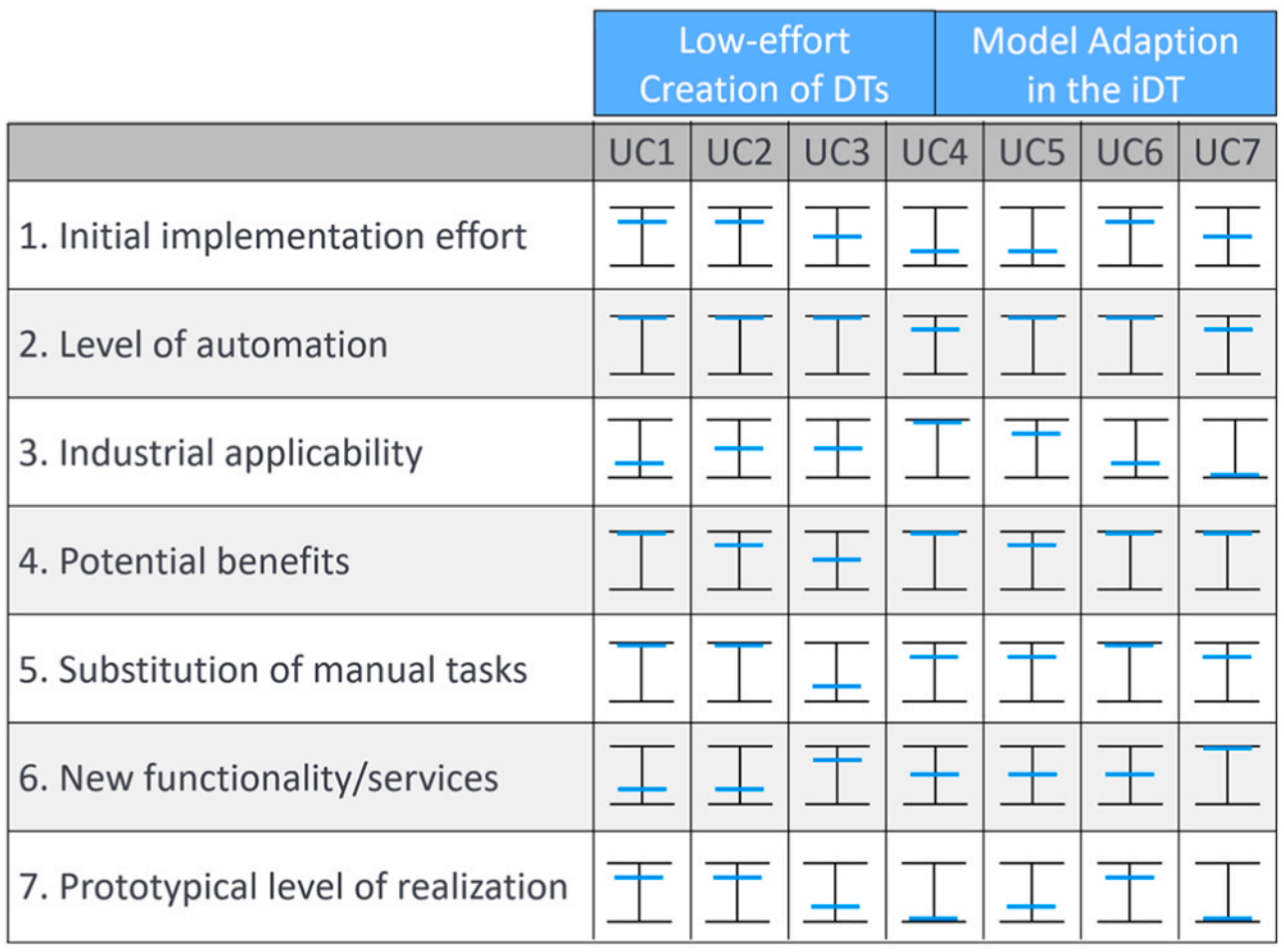

**Fig. 30.** Weighting of the use cases on different criteria.

A closer look at the usefulness of the two concepts in the individual use cases shows that, in principle, it would also be possible to create and use DTs explicitly for the respective use case. However, this would result in significant efforts for the creation. The total effort can simply increase by a factor of five or more. In addition, some benefits can only be realized directly in this way through the consistent use of the DT. For example, instance-specific component data could only be used for more accurate simulations or as a service to increase sensor accuracy, with considerable additional effort during operational phase which would make this use case economically unviable without the DT.

With the presented use cases, the statements made in the literature are substantiated that models are required for different views of a system with abstractions intended for this purpose and that these can also be reused in different phases (Section 2.3). Thereby, in contrast to mostly conceptual proposals of reusing behavior models as White-box, this paper presents a prototypical realization over the entire machine life cycle. Although a very specific, but standard application scenario of discrete manufacturing has been chosen. Nevertheless, the authors think and invite the industry and research community, that due to the proposed model structure according to [26] and the presented use cases, a transfer to other areas can take place. Even in areas such as process technology, different models with different perspectives on the overall system are created [64].

A central challenge for the reuse is above all the provision of information in the form of expert knowledge about the respective models, to adapt and use them. By running through the use cases over the entire life cycle, information such as limit values of variables, calculation times, suitable use cases and model validity ranges were identified in addition to the information resulting from the model structure (modeling depth, modeling width, modeling range). A standardized approach to the challenge is provided by the AAS. But also, with regard to the model coupling of individual components to complete systems, the FMI standard provides a possible realization option. These approaches represent an essential interface between component manufacturers and machine manufacturers as well as machine operators and will be considered in combination with the presented model structure and further above-mentioned Meta-Information in further work.

Another challenge in the efficient creation of DTs is the preparation of the behavior model libraries in all four modeling depths. For this purpose, future work will investigate approaches for the automatic creation of the libraries as well as possibilities for the automatic abstraction of the modeling depths.

In addition to White-box models, the management of black-box models is also of interest for automatic model adaption during the operational phase. These can provide optimal model configurations for some use cases in the sense of the magic triangle, especially due to their fast computing times, e.g. for the pre-simulation of the operation-parallel simulation (Section 4.7) and the prognosis as well as parameter optimization. The appropriate information modeling of these models for interoperable provision must also be considered.

## 5. Conclusion and outlook

This paper presents the combination of efficient behavior model creation and model adaption in an overall concept. The approach presented facilitates the continuous use of behavior models from the Digital Twin along the entire machine lifecycle and across company boundaries in different use cases. In addition, seven industry-relevant use cases are used to illustrate the benefits of behavior models from the Digital Twin along the entire machine lifecycle. Besides that, the contribution shows:

- Challenges in the creation, adaption and use of behavior models arise from the large number of heterogeneous types of behavior models that differ in terms of their modeling approach, simulation tools and simulation properties (solvers, step sizes, …).
- The presented concept for the low-effort creation of Digital Twins in terms of behavior models is an important step to improve the effort-benefit ratio of behavior models over the entire machine life cycle.
- The PDCA method underlying the presented concept of automatic model adaption enables efficient reuse of behavior models beyond the development phase with the help of an agent-based realization.
- The Simulation results obtained from the use cases in the early phases of the machine life cycle provide relevant knowledge transferable via the model comprehension as one key element of the intelligent Digital Twin beneficial for use cases in later phases.
- Based on the different, realized use cases, a guideline for the application of behavior models in the Digital Twin over the machine life cycle is provided for the industry.

Future work will address, among other things, the automatic creation of behavior models in different modeling depths. Furthermore, the topic of interoperable provision of behavior models for automatic model adaption will be investigated. In this context, the necessary expert knowledge about the models and their couplings has to be mapped appropriately.

## Declaration of Competing Interest

The authors declare that they have no known competing financial interests or personal relationships that could have appeared to influence the work reported in this paper.

## Acknowledgements

This contribution was funded by the Federal Ministry of Education and Research (BMBF) under grant 03HY302R.

Furthermore, this work was supported by the Ministry of Science, Research and the Arts of the State of Baden-Wurttemberg within the sustainability support of the projects of the Excellence Initiative II.